\documentclass[onefignum,onetabnum]{siamonline220329}

\newcommand{\specialcell}[2][l]{%
  \begin{tabular}[#1]{@{}c@{}}#2\end{tabular}}
\usepackage{braket,amsfonts}
\usepackage[caption=false]{subfig}
\usepackage{pgfplots}
\usepackage{array}
\usepackage{cite}
\usepackage[utf8]{inputenc}
\usepackage{footnote}
\usepackage{amssymb} 
\usepackage{amsmath}
\usepackage{amsfonts}
\usepackage{multirow}
\usepackage{color}
\usepackage{schemabloc}
\usepackage{bm}
\usepackage{tikz}
\usepackage{footnote}
\usepackage{tabularx}
\usepackage{csquotes}
\usepackage{booktabs} 
\usepackage{stackengine}
\usepackage{makecell}
\usepackage{rotating}
\usepackage[normalem]{ulem}  
\usepackage{multirow}
\usepackage{algorithm}
\usepackage{algpseudocode}
\usetikzlibrary{backgrounds}
\usetikzlibrary{shapes,arrows}
\usepackage{verbatim}
\usepackage{tabu}
\usepackage{xcolor}
\newcommand{\matr}[1]{{\bm{#1}}}
\newcommand{\vect}[1]{{\bm{#1}}}
\newcommand{\norm}[1]{\left\lVert#1\right\rVert}
\newcommand{\prox}{\mathrm{prox}}
\DeclareMathOperator*{\argmax}{argmax} 
\DeclareMathOperator*{\argmin}{argmin}
\usepackage{cleveref}
\usepackage{graphicx,epstopdf} 
\renewcommand{\arraystretch}{1.15}
\Crefformat{equation}{#2Equation~#1#3}
\Crefformat{figure}{#2Figure~#1#3}
\Crefformat{table}{#2Table~#1#3}
\let\eqref\Cref

\newcommand{\rev}[1]{{\color{black} #1}}
\newcommand{\revv}[1]{{\color{black} #1}}

\title{PnP-ReG: Learned Regularizing Gradient for Plug-and-Play Gradient Descent \thanks{ This work was supported by the french ANR research agency in the context of the artificial intelligence project DeepCIM.}}
\author{Rita Fermanian\thanks{Inria Rennes -- Bretagne-Atlantique, 	  
263 Avenue G\'en\'eral Leclerc, 35042 Rennes Cedex, France
(e-mail: firstname.lastname@inria.fr)} \and Mikael Le Pendu\footnotemark[2] \and Christine Guillemot\footnotemark[2]}

\headers{Regularizing Gradient for Plug-and-Play}{Rita Fermanian, Mikael Le Pendu, and Christine Guillemot}

\date{\today}

\begin{document}

\maketitle
\begin{abstract}
The Plug-and-Play (PnP) framework makes it possible to integrate advanced image denoising priors into optimization algorithms, to efficiently solve a variety of image restoration tasks \rev{generally formulated as Maximum A Posteriori (MAP) estimation problems}.
The Plug-and-Play alternating direction method of multipliers (ADMM) and the Regularization by Denoising (RED) algorithms are two examples of such methods that made a breakthrough in image restoration.
\revv{However, the former Plug-and-Play approach only applies to proximal algorithms. And while the explicit regularization in RED can be used in various algorithms, including gradient descent, the gradient of the regularizer computed as a denoising residual leads to several approximations of the underlying image prior in the MAP interpretation of the denoiser.}
We show that it is possible to \rev{train a network directly modeling the gradient of a MAP regularizer while jointly training the corresponding MAP denoiser.}
We use this network in gradient-based optimization methods and obtain better results comparing to other generic Plug-and-Play approaches. We also show that the regularizer can be used as a pre-trained network for unrolled gradient descent. Lastly, we show that the resulting denoiser allows for a better convergence of the Plug-and-Play ADMM.

\end{abstract}



\section{Introduction}
This paper proposes a new approach for solving linear inverse problems in imaging. Inverse problems represent the task of reconstructing an unknown signal from a set of  corrupted observations.
Examples of inverse problems are denoising, super-resolution, deblurring and inpainting, which all are restoration problems.
Supposing that the degradation process is known, we can formulate the restoration task as the minimization of a data fidelity term. However, inverse problems are ill-posed. Hence they do not have a unique solution. A common approach for dealing with this issue consists in introducing prior knowledge on images, in the form of an extra regularization term which penalizes unlikely solutions in the optimization problem.
Early methods 
used handcrafted priors such as total variation (TV) \cite{rudin1992nonlinear,chambolle2010introduction,diptv,admmdiptv} and wavelet regularizers \cite{figueiredo2003algorithm}, as well as low-rank regularizers such as weighted Schatten \cite{lefkimmiatis2013hessian,xie2016weighted} and nuclear \cite{gu2014weighted,gu2017weighted} norms of local image features. 

With the advancement of deep learning, problem-specific deep-neural networks yielded a significant performance improvement in the field of image restoration. In fact, using a reasonable amount of training data, we can train a neural network that can learn a mapping between the space of measurements (i.e. degraded images) and the corresponding solution space (i.e. ground-truth images). These networks interpreted as 
deep regression models are efficient for solving different applications such as sparse signal recovery \cite{mousavi2017learning}, deconvolution and deblurring \cite{xu2014deep,Tan2017,uchida2018non,eboli2020end,NEURIPS2020_0b8aff04}, super-resolution \cite{dong2014learning,Ledig2017,ning2020accurate}, and demosaicing \cite{henz2018deep}. Nevertheless, they have to be designed specifically for each application, hence they lack genericity and interpretability. 

Recent methods have been introduced with the goal of coupling classical optimization with deep learning techniques.
Most of these methods are derived from the idea of the Plug-and-Play prior (PnP) popularized by Venkatakrishnan el al. \cite{PnP1}. The method in \cite{PnP1} uses the ADMM algorithm \cite{boyd2011distributed} which iteratively and alternately solves two sub-problems, each associated to either the regularization term or the data fidelity term. The authors have noted that the regularization sub-problem (formally defined as the proximal operator of the regularization term) can be advantageously replaced by applying a state-of-the art denoiser, hence removing the need for explicitly defining a regularization term. While early works \cite{PnP1, chan2016plug, sreehari2016plug} used traditional denoising methods such as BM3D \cite{dabov2007image} or non-local means \cite{buades2005nlm}, the approach makes it possible to leverage the high performances of deep neural networks for solving various tasks with a single deep denoiser \cite{zhang2017learning, zhang2021plug, rick2017one}.

However, these methods remain limited to proximal algorithms which make use of the proximal operator of the regularization term, hence they do not apply to the simple gradient descent algorithm which instead requires the gradient of the regularization term with respect to the current estimate.
For this reason, despite the growing interest for Plug-and-Play algorithms, as well as the extensive use of gradient descent in machine learning, the Plug-and-Play gradient descent has seldom been studied in the literature.

To the best of our knowledge, only few methods intend to model the gradient of a regularizer for solving inverse problems in such a Plug-and-Play gradient descent scheme: the Regularization by Denoising (RED) proposed by Romano et al. \cite{romano2017little} explicitly defines a regularization term which is proportional to the inner product between the estimated image and its denoising residual obtained with an off-the-shelf denoiser. They show that, under certain conditions, the gradient of this regularization term is the denoising residual itself. However, Reehorst and Schniter \cite{reehorst2018regularization} proved later that the gradient expression proposed in RED is not justified with denoisers that lack Jacobian symmetry, which excludes most practical denoisers such as BM3D \cite{dabov2007image} or state-of-the-art deep neural networks.
\rev{To circumvent that issue, the authors in \cite{hurault2021gradient} and \cite{cohen2021it} plug a denoiser trained to perform an exact gradient step on a regularization function represented by a neural network, leading to convergence guarantees.
However, similarly to RED, the regularization term is explicitly defined by a denoiser. \revv{This involves an additional noise level hyper-parameter that must be tuned per-application, although in theory,} a regularizer (and thus its gradient) should fully determine the image prior, regardless of the task to be solved.}
\revv{
This parameter has an interpretation in the MMSE Bayesian framework.
Given a MMSE Gaussian denoiser for a standard deviation $\sigma$ and a prior distribution $p_{mmse}$, the denoising residual is known to be proportional to the gradient $\nabla[-\ln(p_{mmse}\circ G_\sigma)]$ where $p\circ G_\sigma$ is a ``smoothed'' prior obtained by convolution with a Gaussian of parameter $\sigma$. This motivates the use of a denoiser trained for a small noise standard deviation in \cite{romano2017little, hurault2021gradient, cohen2021it}, to best approximate the ``true'' prior.
However, even if the gradient $\nabla[-\ln(p_{mmse})]$ (called the ``score'') was known perfectly, using it for regularization within a MAP optimization, as in these methods, results in another approximation: as established in \cite{gribonval2011should}, the MMSE denoiser can also be interpreted as a MAP denoiser for some regularizer $-\ln(p_{map})$ (in general $p_{map}\neq p_{mmse}$). A more suitable gradient for solving MAP estimation problems with gradient descent is thus $\nabla[-\ln(p_{map)}]$, which we intend to train with a dedicated network in this paper.

On the other hand, the score $\nabla[-\ln(p_{mmse})]$ is suitable for score-matching methods that solve inverse problems formulated in the MMSE framework \cite{Song2019, Saremi2018, kadkhodaie2020solving, Laumont2022}, hence requiring sampling-based algorithms that are typically slow. Nevertheless, it is worth mentioning the significantly faster Denoising Diffusion Restoration Models (DDRM) \cite{kawar2022denoising} which is a sampling-based solver that leverages diffusion models for restoration purposes.
}

It must also be noted that a network playing the role of the gradient of a regularizer can be trained in the context of the so-called ``unrolled algorithms'' \cite{gregor2010learning, diamond2017unrolled, sun2016deep, schmidt2014shrinkage, mardani2018neural, ning2020accurate, gilton2019neumann, yang2018unrolled, Kobler2020TDV}. Extending on the Plug-and-Play idea, this approach consists in training the regularizing network in an end-to-end fashion so that applying a given number of iterations of the algorithm yields the best results for a specific inverse problem. In particular, the Total Deep Variation (TDV) method \cite{Kobler2020TDV} consists of an unrolled gradient based algorithm where the network represents the regularization function. Due to the end-to-end training, high quality results can be obtained for the targeted application.
\rev{A similar end-to-end training approach has also been exploited in \cite{Adler2017} and \cite{Flynn2019} to learn a network which iteratively updates a current estimate given the gradients of the function to minimize. Here the network represents the algorithm itself rather than the prior.}
However, these end-to-end approaches lose the genericity of Plug-and-Play algorithms. Furthermore, it is not clear what interpretation can be given to the trained network which does not only learn prior knowledge on images, but also task-specific features.

The aim of this paper is then to train a network that models the gradient of a regularizer, without relying on task-specific training.
Hence, our regularizing network can be used for solving inverse problems using a simple gradient descent algorithm, unlike existing Plug-and-Play methods that are suitable only for proximal algorithms.
Our method makes use of a second network pre-trained for the denoising task.
Based on the assumption that the denoiser represents the proximal operator of an underlying differentiable regularizer (defining the image prior), we derive a loss function that links the denoiser and the regularizer's gradient. However, since there is no guarantee that this assumption is mathematically valid for a denoising neural network, we propose an approach where the pre-trained denoiser is modified jointly with the training of our regularizing network.
This approach encourages the denoiser to be consistent with the definition of a proximal operator of a differentiable regularizer, and significantly improves our results in comparison to keeping the denoiser fixed.

We use our network to solve different inverse problems such as super-resolution, deblurring and pixel-wise inpainting in a simple gradient based algorithm, and obtain better results when comparing to other generic methods. We also show that our training method can advantageously serve as a pre-training strategy, later facilitating a per-application tuning of the regularization network in the framework of unrolled gradient descent.

\section{Notations and problem statement} \label{sec:notations}
We consider linear inverse problems which consist in recovering an image $x \in\mathbb{R}^{n}$ from its degraded measurements $y \in\mathbb{R}^{m}$ obtained with the degradation model:
\begin{equation}\label{eq:y}
y=Ax+\epsilon,
\end{equation}
where 
$A\in\mathbb{R}^{m\times n}$ represents the degradation operator depending on the inverse problem and $\epsilon \in\mathbb{R}^{m}$ typically represents Additive White Gaussian Noise (AWGN). 
The restoration of these degraded images is an ill-posed problem, therefore a prior is used to restrict the set of solutions. The reconstruction can be treated using Bayesian estimation that uses the posterior conditional probability $p(x|y)$. Maximum a posteriori probability (MAP) is the most popular estimator in this scheme, where we choose $x$ that maximizes $p(x|y)$.
The estimation task is hence modeled as the optimization problem:

\begin{align}\label{eq:MAP}
       \widehat x_{MAP}&= \argmax_x p(x|y) = \argmax_x \frac{p(y|x) p(x)}{p(y)},
      \\  &=  \argmin_x -log(p(y|x)) - log(p(x)),
\end{align}
where $p(y|x)$ and $p(x)$ are respectively the likelihood and the
prior distributions. For the linear degradation model in \eqref{eq:y} with Additive White Gaussian Noise $\epsilon$ of standard deviation $\sigma$, we get
\begin{equation}\label{eq:MAP2}
       \widehat x_{MAP} =\argmin_x \; \frac{1}{2} \left\| y-Ax \right\|_2 ^2 + \sigma^2 \, \phi(x),
\end{equation}
where the data fidelity term $f(x) = \frac{1}{2} \left\| y-Ax \right\|_2 ^2$ enforces the similarity with the degraded measurements, whereas the regularization term $\phi(x)$ reflects prior knowledge and a property to be satisfied by the searched solution. The non-negative weighting parameter $\sigma^2$ balances the trade-off between the two terms.
The problem in \eqref{eq:MAP2} does not have a closed-form solution in general. Therefore, it must be solved using different optimization algorithms.
The Plug-and-Play framework typically considers proximal splitting algorithms which decompose the problem into two sub-problems (one for each term in \eqref{eq:MAP2}) and solve them alternately. In these algorithms, the regularization sub-problem consists in evaluating the proximal operator of the regularization term defined as:
\begin{equation}
\begin{aligned}\label{eq:prox}
\mathrm{prox}_{\sigma^2\phi}(z) &=\argmin_x \mathcal{F}_\phi(x,z,\sigma),\\
\text{with}\quad\mathcal{F}_\phi (x,z,\sigma) &= \frac{1}{2} \left\| x-z \right\|_2 ^2 +  \sigma^2\,\phi(x).
\end{aligned}
\end{equation}
This can be seen as a particular case of inverse problem where the degradation operator $A$ is the identity matrix, and the degradation only consists in the addition of White Gaussian Noise of standard deviation $\sigma$. The proximal operator in \eqref{eq:prox} can thus be interpreted as a MAP Gaussian denoiser. Hence, this sub-problem can be conveniently replaced by a state-of-the-art Gaussian denoiser in a Plug-and-Play proximal algorithm.

However, this approach does not directly generalize to the gradient descent algorithm where the update formula for the minimization in \eqref{eq:MAP2} is expressed as:
\begin{align}\label{eq:GD1}
\rev{	\widehat x_{k+1} }
    &= \widehat x_{k} - \mu \Big[\nabla f(\widehat x_{k}) + \sigma^2 \nabla \phi(\widehat x_{k})\Big],\\
    &= \widehat x_k - \mu \Big[ A^T ( A \widehat x_k - y) + \sigma^2 \nabla \phi(\widehat x_{k})\Big].
\end{align}

Here, instead of the proximal operator of the regularizer $\phi$, we need its gradient $\nabla \phi$, which cannot be replaced by a denoiser.
In this paper, we propose to train a network $\mathcal{G}$ that can serve as the gradient of the regularization term in Plug-and-Play gradient descent (\cref{alg:PnPGD}).

\rev{
\begin{algorithm}
\caption{\rev{Plug-and-Play Gradient Descent}}\label{alg:PnPGD}
\hspace*{\algorithmicindent} \textbf{Input: } {$y,\sigma, \mu, N$}\\
\hspace*{\algorithmicindent} \textbf{Output: } {$x$}
\begin{algorithmic}[1]
\State $x_0 \gets y$
\For{ $k \gets 0$ to $N-1$}
   \State Apply $G_R \gets \mathcal{G} (x_k)$ 
 \State $x_{k+1} \gets x_k - \mu [ A^T ( A . x_k - y) + \sigma^2 G_R] $
\EndFor
\end{algorithmic}
\end{algorithm}
}

\section{Training of the gradient of a regularizer}
\subsection{Mathematical derivations}
\label{ssec:derivations}
We show in the following that it is mathematically possible to train a network that models the gradient of a regularizer by using a deep denoiser. Let us consider a denoiser $\mathcal{D}_\sigma$ defined as the proximal operator in \eqref{eq:prox}:
\begin{equation}
\label{eq:D_sigma}
\mathcal{D}_\sigma (z) =\argmin_x \mathcal{F}_\phi (x,z,\sigma).
\end{equation}
Hence, for $\sigma$ and $z$ fixed, the denoised image $x=\mathcal{D}_\sigma (z)$ minimizes $\mathcal{F}_\phi (x,z,\sigma)$. Therefore, we have:
\begin{equation}
\label{eq:m0}
\frac{\partial \mathcal{F}_\phi }{\partial x}\Big|_{x= \mathcal{D}_\sigma(z)} = 0,
\end{equation}
Furthermore, $\frac{\partial \mathcal{F}_\phi }{\partial x}$ can be computed as:

\begin{align}
\label{eq:m1} \frac{\partial \mathcal{F}_\phi }{\partial x} &= \frac{\partial \Big[\frac{1}{2} \left\| x-z \right\|_2 ^2 +  \sigma^2\,\phi(x)\Big] }{\partial x},\\
\label{eq:m7}	&=x - z + \sigma^2 . 	\frac{\partial \phi(x)}{\partial x}.
\end{align} 
Evaluating at the denoised image $x=\mathcal{D}_\sigma(z)$ thus gives:
\begin{equation}\label{eq:m2}
	\frac{\partial \mathcal{F}_\phi }{\partial x}\Big|_{x= \mathcal{D}_\sigma(z)} = \mathcal{D}_\sigma(z) - z +  \sigma^2 . 	\frac{\partial \phi(x)}{\partial x}\Big|_{ x= \mathcal{D}_\sigma(z)}.
\end{equation} 
Using \eqref{eq:m0} and \eqref{eq:m2}, we obtain:
\begin{align}
\label{eq:m4} 	 \mathcal{D}_\sigma(z) - z +  \sigma^2 . 	\frac{\partial \phi(x)}{\partial x}\Big|_{ x= \mathcal{D}_\sigma(z)} &=0,\\
\label{eq:m5} \sigma^2 .	\frac{\partial \phi(x)}{\partial x}\Big|_{ x= \mathcal{D}_\sigma(z)} &=  z-\mathcal{D}_\sigma(z),\\
\label{eq:m6} \sigma^2. \nabla\phi(\mathcal{D}_\sigma(z)) &= z-\mathcal{D}_\sigma(z).
\end{align} 
Using \eqref{eq:m6}, we can thus train a network \rev{$\mathcal{G}$} to model $\nabla\phi$ (i.e. gradient of the regularizer with respect to its input image) using the loss function:
\begin{equation}\label{eq:LR}
	\mathcal{L}_{\rev{\mathcal{G}}} = \Bigg\|\sigma^2 \big[ \rev{\mathcal{G}}(\mathcal{D}_\sigma(z)) \big]- (z-\mathcal{D}_\sigma(z)) \Bigg\|_2 ^2.
\end{equation} 
This requires the knowledge of the corresponding denoiser $\mathcal{D}_\sigma$.
Note that \eqref{eq:m6} is valid for any value of $\sigma$ regardless of the degradation in $z$. Hence, $\sigma$ can be seen as a free parameter of our loss $\mathcal{L}_{\rev{\mathcal{G}}}$. For small values of $\sigma$, the input $\mathcal{D}_\sigma(z)$ of the network \rev{$\mathcal{G}$} will be close to the degraded image $z$. Hence \rev{$\mathcal{G}$} will be trained to fit the artifacts in the degraded images (e.g. noise). On the other hand, for high values of $\sigma$, the input of \rev{$\mathcal{G}$} will be a strongly denoised image, with reduced artifacts but less details. Hence, \rev{$\mathcal{G}$} will be trained to recover the missing details. During the training, we vary the value of this parameter so that our network can recover details while also removing artifacts (see details in \Cref{ssec:train_framework}).

Also note that in practice, since $\phi$ is meant to be used in gradient-based algorithms, we only need the gradient $\nabla\phi$ rather than an explicit definition of $\phi$. Hence, we propose in what follows a framework for training \rev{$\mathcal{G}$} jointly with the denoiser $\mathcal{D}$, \rev{so that $\mathcal{G}$ can then be used in place of $\nabla\phi$ in Plug-and-Play gradient descent. In the remainder of the paper, we refer to $\mathcal{G}$ as the regularizing gradient network (ReG).}

\tikzstyle{block} = [draw, rectangle, 
    minimum height=3em, minimum width=6em]
\tikzstyle{sum} = [draw, circle, node distance=1cm]
\tikzstyle{input} = [coordinate]
\tikzstyle{output} = [coordinate]
\tikzstyle{pinstyle} = [pin edge={to-,thin,black}]

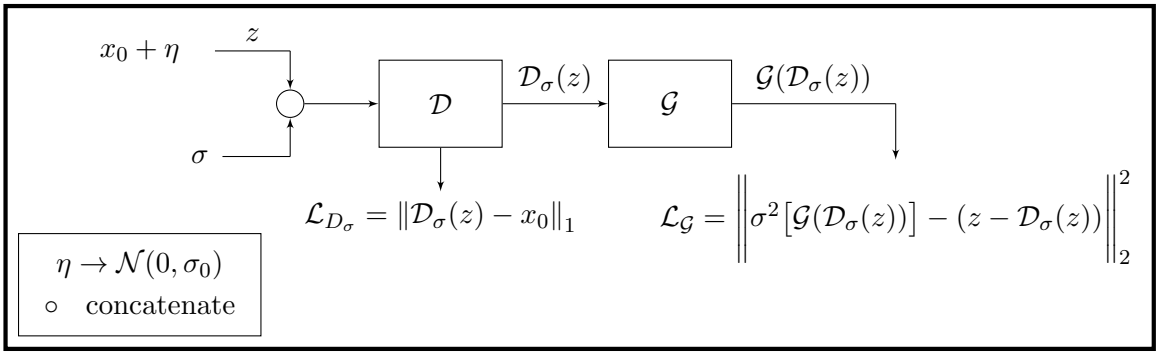
\begin{figure*}[ht]
        \centering
       
      \begin{tikzpicture}[auto, node distance=2cm,>=latex',framed,background rectangle/.style={ultra thick,draw=black}]
    \node [name=in, node distance= 0cm]{$x_0+\eta$};
    \node [input, name=input,right of =in,node distance= 1cm]  {};
    \node [input, name=a1, right of=input,node distance=1cm] {};
    \node [sum,  below of=a1,node distance=0.7cm] (sum) {};
    \node [input, name=a2, below of=sum,node distance=0.7cm] {};
    \node [input, name=input2,  left of=a2,node distance=0.9cm] {};
    \node[name=in, left of=input2,node distance = 0.3cm]{$\sigma$};
    \node [block, name= denoiser, right of=sum,node distance=2cm, minimum width=4.25em] (controller) {$\mathcal{D}$};
    \node [block, name = regularizer, right of=controller, node distance=3.05cm, minimum width=4.25em] (system) {\rev{$\mathcal{G}$}};
    \node [input, name=a3, right of=system,node distance=3cm]{};

    \node [name=l2, below of=a3,node distance=1.5cm] {$\mathcal{L}_{\rev{\mathcal{G}}} = \Bigg\|\sigma^2 \big[\rev{\mathcal{G}}(\mathcal{D}_\sigma(z)) \big]- (z-\mathcal{D}_\sigma(z)) \Bigg\|_2^2$};
    \node [name=l1, below of=controller,node distance=1.5cm] {$\mathcal{L}_{D_\sigma} = \left\| \mathcal{D}_\sigma(z)- x_0 \right\|_1 $};
    \node at (0,-3.1) [block, name= legend] (legend) {\begin{tabular}{c} $\eta \rightarrow \mathcal{N}(0,\sigma_0)$\\ $\circ$ \; concatenate \end{tabular}};
    \draw [draw,-] (input) -- node {$z$} (a1);
    \draw [draw,->] (a1) -- node {} (sum);
    \draw [draw,-] (input2) -- node {} (a2);
    \draw [draw,->] (a2) -- node {} (sum);
    \draw [->] (sum) -- node {} (controller);
    \draw [->] (controller) -- node{$\mathcal{D}_\sigma(z)$} (system);
    \draw [draw,-] (system) -- node {$\rev{\mathcal{G}} (\mathcal{D}_\sigma(z))$} (a3);
    \draw [draw,->] (a3) -- node {} (l2);
    \draw [draw,->] (controller) -- node {} (l1);
\end{tikzpicture}
        
\caption{\small{Framework for joint training of $ \mathcal{D}$ and \rev{$\mathcal{G}$}.}}
\label{fig:framework}
\end{figure*}

\subsection{Training framework for the regularizing gradient network $\mathcal{G}$}
\label{ssec:train_framework}
The training framework is depicted in \Cref{fig:framework}.
Let \mbox{$\eta \rightarrow \mathcal{N}(0,\sigma_0)$} be a white Gaussian noise of mean $0$ and standard deviation $\sigma_0$ that we use to corrupt the ground truth images $x_0$ of the training dataset to produce degraded images $z=x_0+\eta$. Let $\sigma$ be a standard deviation value used as a parameter of our loss $\mathcal{L}_{\rev{\mathcal{G}}}$, as defined in \Cref{ssec:derivations}. In order to handle different values of $\sigma$ in \eqref{eq:LR}, $\mathcal{D}_\sigma$ is modeled as a non-blind deep denoiser that takes as input a noise level map (i.e. each pixel of the noise level map being equal to $\sigma$) concatenated with the noisy image $z$. \revv{This approach has been previously suggested in \cite{zhang2018ffdnet,al2020learning,zhang2021plug}}.
\rev{Note that if the denoiser $D_\sigma$ does not satisfy the formal definition of a MAP Gaussian denoiser for some differentiable prior, there may not exist a function $\nabla\phi$ that satisfies \eqref{eq:m6}. We prevent this issue by starting from a pre-trained denoiser which we jointly update along with the training of the ReG network $\mathcal{G}$. In order to preserve the denoising performance of $D_\sigma$ during the training, we use an additional denoising loss $\mathcal{L}_{D_\sigma}$ defined as:}
\begin{equation}\label{eq:L1}
	\mathcal{L}_{D_\sigma} = \left\| \mathcal{D}_\sigma(z)- x_0 \right\|_1.
\end{equation} 
Note that \eqref{eq:L1} is a suitable loss for the denoiser only when $\sigma=\sigma_0$ since the non-blind denoiser must be parameterized with the true noise level $\sigma_0$ of the noisy input.
The denoised output $\mathcal{D}_\sigma(z)$ is then inputted to the network modeling the gradient of the regularizer in order to train it using the loss $\mathcal{L}_{\rev{\mathcal{G}}}$ defined in \eqref{eq:LR}.

Hence, our goal is to minimize the global loss defined as:
\begin{equation}\label{eq:L}
	\mathcal{L} = \delta \; \mathcal{L}_{\mathcal{D}_\sigma} + \lambda \;  \mathcal{L}_{\rev{\mathcal{G}}}, 
\end{equation} 
where $\lambda > 0 $ and 
$
\delta = \left\{
    \begin{array}{ll}
        1 &\mbox{if  } \sigma_0 = \sigma \\
        0 & \mbox{otherwise}
    \end{array}
\right.
$.\\

For training the deep denoiser network, we should set the noise level $\sigma$ inputted to the network equal to the actual noise level $\sigma_0$ used for generating $\eta$. However, as explained in \Cref{ssec:derivations}, for the loss $\mathcal{L}_{\rev{\mathcal{G}}}$, it is preferable to select $\sigma$ independently of $\sigma_0$. Hence, the input $\mathcal{D}_\sigma(z)$ of the ReG network \rev{$\mathcal{G}$} can cover a wide range of alterations, including images with remaining noise (i.e. $\sigma<\sigma_0$) or with too strong denoising, and thus less details (i.e. $\sigma>\sigma_0$).
We therefore choose to alternate during the training between either selecting independently $\sigma$ and $\sigma_0$, or setting $\sigma = \sigma_0$ in order to keep $\mathcal{D}$ faithful to the data. Furthermore, since the denoiser loss $\mathcal{L}_{\mathcal{D}_\sigma}$ is only valid when $\sigma = \sigma_0$, we omit this loss when $\sigma \neq \sigma_0$ by setting $\delta=0$.

\rev{
Note that we only need to alternate between setting $\sigma=\sigma_0$ and $\sigma\neq\sigma_0$ because we choose to train the denoiser jointly with the ReG network. A simpler training strategy would consist in first training the denoiser separately (only with the loss $\mathcal{L}_{D_\sigma}$ and with $\sigma=\sigma_0$), and then training the regularizer (only with the loss $\mathcal{L}_{\rev{\mathcal{G}}}$ and with $\sigma\neq\sigma_0$).
However, our experimental results in \Cref{ssec:joint_train} clearly show the advantage of our joint training scheme, which confirms that the separately trained denoiser does not satisfy the assumption of a Gaussian MAP denoiser for a differentiable prior.
}

\begin{figure*}
    \centering
    \includegraphics[width=\textwidth]{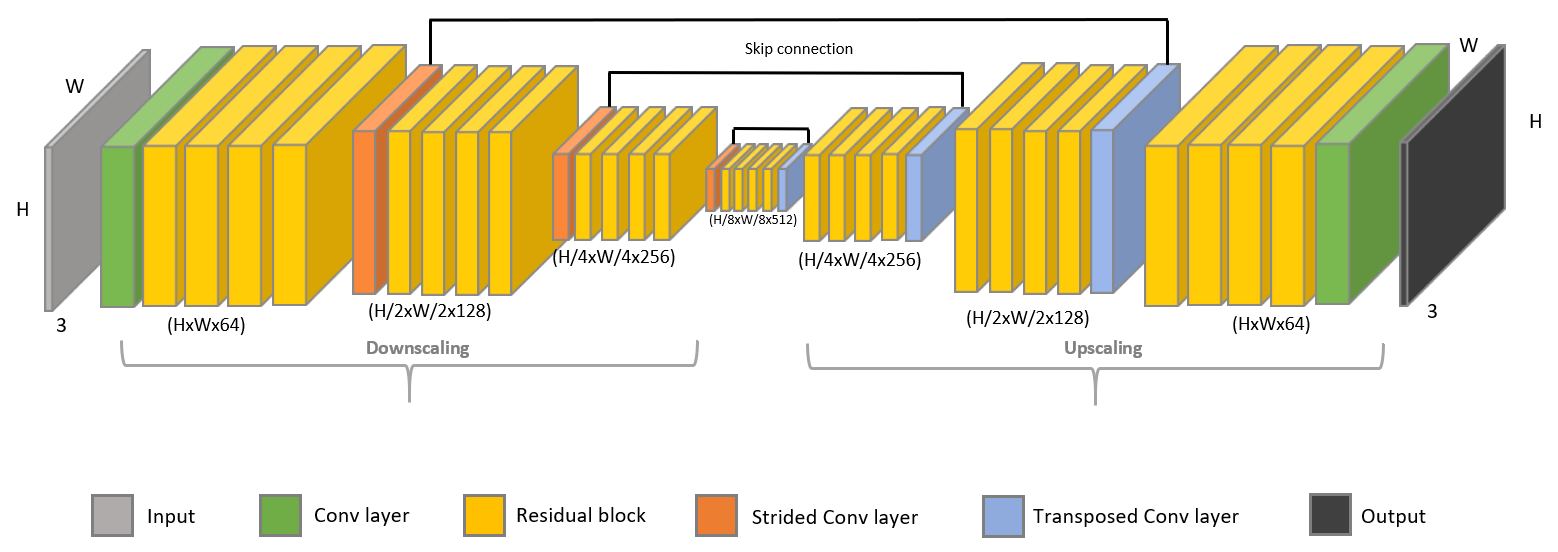}
    \caption{\small{Architecture of the DRUNet network \cite{zhang2021plug}, that we have chosen for \rev{$\mathcal{G}$} by changing the input channel to 3 instead of 4 (the ReG network does not need a noise level map as additional input as it does not depend on noise level).}}
\label{fig:architecture}
\end{figure*}
\subsection{Training details}
For the training, we have used a state-of-the-art deep denoiser architecture in order to train our ReG network. We have chosen to work with the DRUNet proposed in \cite{zhang2021plug} which is a combination of U-Net \cite{ronneberger2015u} and ResNet  \cite{he2016deep}. \rev{DRUNet additionally uses a bias-free network architecture, which has been shown to allow for good generalization of denoisers over various noise levels, even if they were not seen during training \cite{mohan2020robust}.}
Since it takes as input the noisy image concatenated in the channel dimension with a noise level map, it can suitably represent the non-blind denoiser $\mathcal{D}_\sigma$.

The architecture of \rev{$\mathcal{G}$} is shown in \Cref{fig:architecture}. It is the same architecture as the \rev{bias-free} DRUNet denoiser, with the only difference that it does not take a noise level map as additional input.

We have initialized $\mathcal{D}_\sigma$ using the pre-trained DRUNet denoiser (which we have reproduced based on the work in \cite{zhang2021plug}). Then, we have trained \rev{$\mathcal{G}$} while jointly updating $\mathcal{D}_\sigma$, following the proposed framework in \Cref{ssec:train_framework}. The weight $\lambda$ of the loss $\mathcal{L}_{\rev{\mathcal{G}}}$ in \eqref{eq:L} has been set equal to $0.004$. The parameters $\sigma$ and $\sigma_0$ have been selected following the alternating strategy described in \Cref{ssec:train_framework}: for half of the training iterations, we have used \mbox{$\sigma=\sigma_0$} with a value chosen randomly with uniform distribution in $[0,50]$; otherwise, $\sigma$ and $\sigma_0$ have been chosen independently with the same uniform distribution.

The remaining training details are similar to the ones presented in \cite{zhang2021plug} for the DRUNet pre-training:
the same large dataset of 8694 images composed of images from the Waterloo Exploration Database \cite{ma2016waterloo}, the Berkeley Segmentation Database \cite{chen2016trainable}, the DIV2K dataset \cite{agustsson2017ntire} and the Flick2K dataset \cite{lim2017enhanced} has been used. 16 patches of 128x128 have been randomly sampled from the training dataset for each iteration. We have used the ADAM optimizer \cite{kingma2014adam} to minimize the loss $\mathcal{L}$ defined in \eqref{eq:L}. The learning rate has been initially set to 1e-4, and decreased by half every 100,000 iterations until reaching 5e-7, where the training stops.

\section{Experimental results}

\vspace{5pt}

\subsection{Plug-and-Play gradient descent}
\rev{In this section, we evaluate the performance of the Plug-and-Play gradient descent based on our ReG network. We refer to this approach as Plug-and-Play Regularizing Gradient (PnP-ReG). We perform the evaluation on several inverse problems: super-resolution, deblurring and pixel-wise inpainting. Evaluations in this section are performed over the Set5 \cite{Set5} and the CBSD68 \cite{CBSD68} test datasets.}

As the main goal of this approach is to solve inverse problems using simple gradient-based algorithms with a generic regularizer, we compare ourselves to algorithms that are designed to solve different inverse problems using a single regularization network in a Plug-and-Play framework. Hence, we compare the performance of our PnP-ReG method to the PnP-ADMM with the DRUNet denoiser from \cite{zhang2021plug} \rev{(see Appendix \ref{sec:pnpadmm} for the classical ADMM formulation and parameterization)}; the RED method \cite{romano2017little} in gradient descent with the same DRUNet denoiser; \rev{GS-PnP \cite{hurault2021gradient} where the network architecture used in the denoising function is the DRUNet as well;} and Chang's projection operator (One-Net) \cite{rick2017one} used in an ADMM framework. 
\rev{We also include a comparison with the Implicit Prior of \cite{kadkhodaie2020solving} only for pixel-wise inpainting, since the method is based on the assumption that the kernel matrix is
semi-orthogonal, which is not the case for the super-resolution and deblurring applications in our work. }

For fair comparisons, we have reproduced all the results under the same conditions, i.e. using the same initialization and the same degradation operator $A$ as described in the following subsections for each application. We have tuned the parameters of each method \rev{on the Set5 dataset} for each application to obtain the best results, \rev{and we have used the same parameters for the CBSD68 dataset.}



\subsubsection{Parameters setting and implementation details}
\label{ssec:ParamSettings}
 \rev{In this section, we give further implementation and parameterization details for our method as well as for the reference methods.}

\rev{For the experimental results, we have used the ADAM optimizer \cite{kingma2014adam} instead of the simple gradient step (i.e. line 4 in \cref{alg:PnPGD}) to solve \eqref{eq:MAP2} in the Plug-and-Play framework. Similarly to \cite{zhang2021plug}, we have applied a periodical geometric self-ensemble data-augmentation method. This involves one transformation (e.g. flipping, rotation) on the input of the network and the counterpart inverse transformation on the output.}
\Cref{tab:Parameters PnP} shows the parameters used during testing for \rev{PnP-ReG. The number of iterations $N$ is set to 1500.} In theory, the parameter $\sigma$ in \eqref{eq:MAP2} should be equal to the true standard deviation $\sigma_n$ of the Gaussian noise added on the degraded image.
However, when $\sigma_n = 0 $ (e.g. \rev{noiseless} super-resolution, pixel-wise inpainting), choosing $\sigma = 0 $ would completely remove the regularization term. For these cases, we choose a small non-zero value of $\sigma$ depending on the application.

To reproduce the results of \cite{rick2017one} we have used the model trained by the authors, which takes input images of size 64x64. Hence we have applied the network on quarter-overlapping sample patches in order to enhance the results by avoiding block artifacts. \rev{Table \ref{tab:Parameters OneNet} in Appendix \ref{sec:parameters} lists the tuned hyper-parameters for the reproduction of \cite{rick2017one}.}

\rev{We have implemented RED \cite{romano2017little} with Gradient Descent using the DRUNet denoiser. For fair comparisons with our method, we have used the ADAM optimizer to perform the gradient update step.
The number of iterations is set to 300 for the tasks of Super-Resolution and Deblurring, and 800 in the case of pixel-wise inpainting. Table \ref{tab:Parameters RED} in Appendix \ref{sec:parameters} shows the tuned parameters for the implementation of RED. 

For GS-PnP \cite{hurault2021gradient}, we have used the parameters described in the paper, except for the standard deviation parameter of the denoiser in Super-Resolution (noiseless or low-noise cases) and pixel-wise inpainting. In general, for a noise standard deviation $\sigma_n$ in the degradation, the authors of \cite{hurault2021gradient} suggest to parameterize the denoiser with a standard deviation of $2*\sigma_n$. However, this is not suitable for the noiseless applications where $\sigma_n=0$ (e.g. noiseless super-resolution, pixel-wise inpainting). Therefore, we have tuned this parameter and have set a denoiser standard deviation of $6/255$ for Super-Resolution in our paper. For pixel-wise inpainting, we have set this parameter to $50/255$ for the first 20 iterations and to $20/255$ for subsequent iterations.
The remaining parameters were chosen as described in \cite{hurault2021gradient}.

For the pixel-wise inpainting results of the Implicit Prior in \cite{kadkhodaie2020solving}, we have used the parameters suggested by the authors in the paper. We have computed the results by averaging 10 samples obtained with their stochastic method, as done in the original paper.

Finally, we have compared with the PnP-ADMM using DRUNet \cite{zhang2021plug}. The PnP-ADMM formulation and its parameterization is elaborated in Appendix \ref{sec:pnpadmm}.
Furthermore, we have applied the periodical geometric self-ensemble data augmentation during the test as done in \cite{zhang2021plug}.
\cref{tab:Parameters PnPADMM} in Appendix \ref{sec:parameters} shows the parameters that yielded the best results for the PnP-ADMM. We note that in the cases where we have added noise, fixing the standard deviation parameter of the denoiser gave better results than decreasing it throughout the iterations.

While in our experiments, we have fine-tuned the parameters of the different methods, including PnP-ADMM, an alternative strategy such as proposed in \cite{wei2020tuning} based on reinforcement learning could be envisaged for automatic search of the algorithm's parameters. 


}



{\renewcommand{\arraystretch}{0.9}
\begin{table}[h]
\caption{{\footnotesize{Parameters used for our PnP-ReG method. $\mu$: gradient step size, $\sigma$: weight of the regularization, $\sigma_n$: standard deviation of the AWGN added on the degraded image.}}}
\setlength\tabcolsep{4pt} 
\vspace{5pt}
    \centering
    {\normalsize
        \begin{tabular}{l l  c| c c c c}
            \toprule
            && $\sigma_n*255$ & \hspace*{0.5cm}& $\mu$ & $\sigma*255$   \\ 
            \midrule
            \multirow{6}{*}{Super-Resolution x2} 
            & \multirow{3}{*}{Bicubic} 
             & 0 && 0.008 & 1.2 \\ 
              && $\sqrt{2}$ && 0.002 & $\sqrt{2}$ \\ 
              && 2.55 && 0.002 & 2.55 \\
             \cmidrule{2-6}
    &\multirow{3}{*}{Gaussian}                  
            & 0 & &0.008 & 0.8 \\ 
              && $\sqrt{2}$ & & 0.002 & $\sqrt{2}$ \\ 
              && 2.55 & & 0.002 & 2.55 \\
    \midrule
     \multirow{6}{*}{Super-Resolution x3} & 
                \multirow{3}{*}{Bicubic} 
            & 0 & &0.002 & 0.9 \\ 
              && $\sqrt{2}$ & &0.002 & $\sqrt{2}$ \\ 
              && 2.55 & &0.002 & 2.55 \\
             \cmidrule{2-6}
    &\multirow{3}{*}{Gaussian}     
    & 0 && 0.004 & 0.4 \\ 
              && $\sqrt{2}$ && 0.002 & $\sqrt{2}$ \\ 
              && 2.55 & &0.002 & 2.55 \\
    \midrule
         \multirow{6}{*}{Deblurring} & 
                \multirow{3}{*}{\includegraphics[width=0.05 \linewidth]{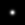}  \includegraphics[width=0.051\linewidth]{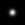}} 
             & $\sqrt{2}$ & &0.004 & $\sqrt{2}$ \\ 
              && 2.55 & &0.004 & 2.55 \\ 
              && 7.65 && 0.004 & 7.65 \\
             \cmidrule{2-6}
                   & \multirow{3}{*}{\includegraphics[width=0.05 \linewidth]{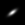}  \includegraphics[width=0.05\linewidth]{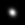}} 
             & $\sqrt{2}$ && 0.005 & $\sqrt{2}$ \\ 
              && 2.55 && 0.005 & 2.55 \\ 
              && 7.65 && 0.005 & 7.65 \\

    \midrule
    \multirow{2}{*}{Pixel-wise inpainting }
    
    &$0.1$       & 0& &0.025 & 3.6/255 \\
& $0.2$ & 0 &&0.01 & 1/255 \\ 
           
            \bottomrule
        \end{tabular}
    }
    
    \label{tab:Parameters PnP}
    \vspace{-8pt}
\end{table}}
 
\vspace{7pt}

\subsubsection{Super-Resolution}~\\ 
Super-resolution consists of reconstructing a high-resolution image from a low-resolution (i.e. downsampled) measurement. Low resolution images have been generated by applying a convolution kernel followed by a downsampling by a factor $t$. We have evaluated our method with bicubic and Gaussian convolutional kernels, with both 2x and 3x downsampling scales and \rev{Gaussian noise with 3 different noise levels $\sigma_n=\{0.0,\sqrt{2}$, 2.55\}/255.} The Gaussian kernel has a standard deviation of $\sigma_b = 0.5\cdot t$ (i.e. $\sigma_b=1$ for x2 and $\sigma_b=1.5$ for x3).  In all the cases, the gradient descent has been initialized with a high resolution image obtained by the bicubic upsampling of the degraded image.

\begin{figure}
\centering
\setlength\tabcolsep{1pt}

\begin{tabular}{c c c }
            Ground Truth   & One-Net   & GS-PnP \\
            \includegraphics[width=0.23\linewidth]{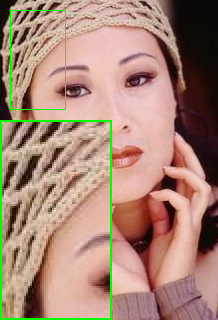} &
            \includegraphics[width=0.23\linewidth]{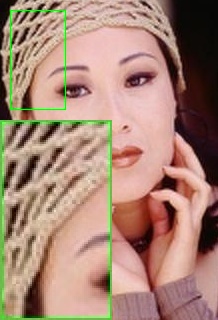} &
            \includegraphics[width=0.23\linewidth]{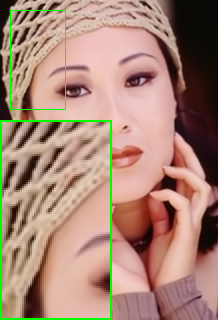}  \\
            &  32.50 dB & 34.33 dB \\
        \end{tabular}

\centering
\begin{tabular}{c c c}
            RED   & PnP-ADMM  & \rev{PnP-ReG} (Ours) \\
            \includegraphics[width=0.23\linewidth]{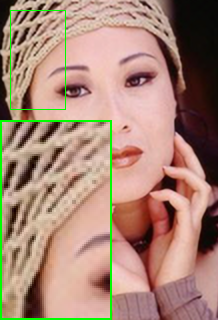} &
            \includegraphics[width=0.23\linewidth]{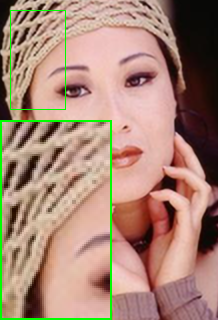} &
            \includegraphics[width=0.23\linewidth]{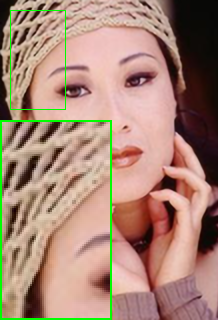} 
             \\
              34.58 dB & 34.34 dB & 34.87 dB \\
        \end{tabular}

  \caption{\small{Visual comparison of super-resolution results obtained with the projection operator (One-Net) \cite{rick2017one}, \rev{ GS-PnP \cite{hurault2021gradient},} RED \cite{romano2017little}, PnP-ADMM \rev{with the denoiser of} \cite{zhang2021plug} and our PnP-ReG method. Low resolution images generated with a bicubic kernel followed by a downsampling by a factor of 2.}}
  \label{fig:PnP-SR2B}
  \vspace{-10pt}

\end{figure}

\Cref{tab:SR2 PnP,tab:SR3 PnP} give a numerical comparison of our method with the aforementioned generic approaches for super-resolution of factor 2 and 3 respectively, for both bicubic and Gaussian filters. The PSNR (peak signal-to-noise ratio) measures presented in this paper are computed on the RGB channels. Numerical comparison gives higher values for PnP-ReG compared to the existing generic Plug-and-Play approaches.
\Cref{fig:PnP-SR2B} shows a visual comparison of the results for a noiseless degradation with a bicubic kernel and a downsampling by a factor of 2. We observe sharper images with less aliasing artifacts produced by our approach.
{\renewcommand{\arraystretch}{0.9}

 \begin{table*}[h]
\caption{\footnotesize{{ Super-resolution results (measured in PSNR [dB]) obtained with our PnP-ReG method; the input images have been corrupted using bicubic and Gaussian kernels followed by downsampling by a factor of 2 \rev{ and adding Gaussian noise with 3 different noise levels $\sigma_n=\{0.0,\sqrt{2}$, 2.55\}/255.}  Comparison with the projection operator (One-Net) \cite{rick2017one}, \rev{  GS-PnP \cite{hurault2021gradient},} RED \cite{romano2017little} and the PnP-ADMM \rev{with the denoiser of} \cite{zhang2021plug}.}}}
 \vspace{5pt}
\setlength\tabcolsep{5pt} 
    \centering
    { \normalsize
        \begin{tabu}{l l  c c c c c c c }
            \toprule[1.2pt]
            & &  \multicolumn{3}{c}{(i) Bicubic } & & \multicolumn{3}{c}{(ii) Gaussian}  \\
            \cmidrule{3-5} \cmidrule{7-9} 
             &      $\sigma_n*255$         &  \footnotesize{$0.0$}       &
             \footnotesize{{$\sqrt{2}$}}   & \footnotesize{$2.55$}  &    &
             \footnotesize{$0.0$} &
             \footnotesize{{$\sqrt{2}$ }}  & \footnotesize{$2.55$}     \\
            \midrule
           \multirow{5}{*}{ Set5}          
               &  OneNet&     33.22  & 33.70 & 33.20 &        & 32.67& 33.08 &  32.45          \\
           
           &  GS-PnP&     34.58  & 34.49 & \textbf{34.31} &        & 33.98& 33.88&  \textbf{33.59}          \\
           &  RED&     35.05 & 34.49 & 33.78 &        & 34.99& 33.80 &  32.84          \\
&  PnP-ADMM&     35.20  & 34.42 & 33.80 &        & 35.14& 33.69 &  32.74         \\

  &  PnP-ReG (Ours)&     \textbf{35.34}  & \textbf{34.90} & 34.29 &        & \textbf{35.30} & \textbf{34.33} &  33.41         \\
            \midrule
            
           \multirow{5}{*}{CBSD68}        
           
           &  OneNet&     29.48  & 29.80 & 29.56 &        & 29.08 & 29.48&  29.06          \\
           
           &  GS-PnP&     29.95  & 29.93 & 29.81 &        & 29.57& 29.51&  29.39          \\
           
           &  RED&     30.45 & 30.14 & 29.73 &        & 30.39& 29.61 & 28.99          \\
           
&  PnP-ADMM&     30.48  & 30.17 & 29.92 &        & 30.42& 29.78 & 29.23         \\

   & PnP-ReG (Ours)&     \textbf{30.55}  & \textbf{30.34} & \textbf{30.06} &        & \textbf{30.50} & \textbf{30.01} &  \textbf{29.44}        \\
            \bottomrule[1.2pt]
        \end{tabu}
    }
    
    \label{tab:SR2 PnP}
    \vspace{-8pt}
\end{table*}}


{\renewcommand{\arraystretch}{0.9}

 \begin{table*}[h]
\caption{\footnotesize{{ Super-resolution results (measured in PSNR [dB]) obtained with our PnP-ReG method; the input images have been corrupted using bicubic and Gaussian kernels followed by downsampling by a factor of 3 \rev{ and adding Gaussian noise with 3 different noise levels $\sigma_n=\{0.0,\sqrt{2}$, 2.55\}/255.}  Comparison with the projection operator (One-Net) \cite{rick2017one}, \rev{ GS-PnP \cite{hurault2021gradient},}  RED \cite{romano2017little} and the PnP-ADMM \rev{with the denoiser of} \cite{zhang2021plug}.}}}
 \vspace{5pt}
\setlength\tabcolsep{5pt} 
    \centering
    { \normalsize
        \begin{tabu}{l l c  c  c c c c c }
            \toprule[1.2pt]
            & &  \multicolumn{3}{c}{(i) Bicubic } & & \multicolumn{3}{c}{(ii) Gaussian}  \\
            \cmidrule{3-5} \cmidrule{7-9} 
           &      $\sigma_n*255$         &  \footnotesize{$0.0$}       &
             \footnotesize{$\sqrt{2}$}   & \footnotesize{$2.55$}  &    &
             \footnotesize{$0.0$} &
             \footnotesize{$\sqrt{2}$ }  & \footnotesize{$2.55$}     \\
            \midrule
           \multirow{5}{*}{ Set5}     
           &  OneNet&     29.65  & 30.18 & 29.79 &        & 29.52& 29.71& 29.05          \\
         &  GS-PnP&     31.48 & 31.43 & \textbf{31.24} &        & 30.91& 30.84& \textbf{30.59}          \\
           &  RED&     31.47 & 31.22 & 30.78 &        & 31.44& 30.77 & 30.05          \\
&  PnP-ADMM&     31.49  & 31.05 & 30.39 &        & 31.45& 30.22 & 29.17         \\

 & PnP-ReG (Ours)&     \textbf{31.75}  & \textbf{31.46} & 31.13 &        & \textbf{31.60} & \textbf{31.09} &  30.39        \\
            \midrule
           \multirow{5}{*}{CBSD68}      
         &  OneNet&     26.17  & 26.89 & 26.67 &        & 25.21& 26.69&  26.33          \\
         &  GS-PnP&     27.11  & 27.08 &26.99 &        & 26.80 & 26.73& 26.64          \\
           &  RED&     27.50 & 27.33 & 27.11 &        & 27.40& 27.06 & 26.24          \\
&  PnP-ADMM&     27.51  & 27.33 & 26.97 &        & \textbf{27.47}& 26.83 & 26.20         \\

  &   \rev{PnP-ReG (Ours)}&     \textbf{27.55}  & \textbf{27.41} & \textbf{27.23} &        & 27.46 & \textbf{27.17}& \textbf{26.77}        \\

            \bottomrule[1.2pt]
        \end{tabu}
    }
    
    \label{tab:SR3 PnP}
    \vspace{-8pt}
\end{table*}}

\vspace{5pt}

\subsubsection{Deblurring}~\\ 
For image deblurring, the degradation consists of a convolution performed with circular boundary conditions.
\rev{We have degraded our images using two 25x25 isotropic Gaussian blur kernels of standard deviations of 1.6 and 2.0, as well as two anisotropic kernels that have been used in \cite{zhang2020deep}. We have considered White Gaussian noise with 3 noise levels $\sigma_n=\{\sqrt{2}$, 2.55, 7.65\}/255. The blurred image is directly used as the initialization of the Plug-and-Play gradient descent.

\Cref{tab:Deblurring PnP} gives the PSNR results [dB] obtained with our method when deblurring images which have been degraded with isotropic and anisotropic kernels respectively. We observe higher PSNR values with respect to the other generic methods for most cases when the added noise levels are $\sigma_n=\{\sqrt{2}, 2.55\}/255$. However, by increasing the noise level to $7.65/255$, the best results have been obtained with GS-PnP. The visual comparison in \Cref{fig:PnP-DK2} shows that our approach successfully recovers the details without increasing the noise.

}

\vspace{5pt}

\vspace{5pt}
{\renewcommand{\arraystretch}{0.9}

 \begin{table*}[h]
\caption{\footnotesize{{Deblurring results (measured in PSNR [dB]) obtained with our PnP-ReG method. The input blurred images have been generated using two isotropic Gaussian kernels of respective standard deviations 1.6 and 2.0, and two anisotropic Gaussian kernels from \cite{zhang2021plug} \rev{ followed by adding Gaussian noise with 3 different noise levels $\sigma_n=\{\sqrt{2}$, 2.55,7.65\}/255.} We compare with the projection operator (One-Net) \cite{rick2017one}, \rev{GS-PnP \cite{hurault2021gradient},}  RED \cite{romano2017little} and the PnP-ADMM \rev{with the denoiser of} \cite{zhang2021plug}.}}}
 \vspace{5pt}
\setlength\tabcolsep{5pt} 
    \centering
    { \normalsize
        \begin{tabu}{l l c c c c c c c}
            \toprule[1.2pt]
            & &  \multicolumn{3}{c}{\includegraphics[width=0.08\linewidth]{kernel2.PNG} } & & \multicolumn{3}{c}{
   \includegraphics[width=0.08\linewidth]{kernel3.PNG} }  \\
            \cmidrule{3-5} \cmidrule{7-9} 
            &      $\sigma_n*255$         &  \footnotesize{$\sqrt{2}$}   & \footnotesize{$2.55$}  &   \footnotesize{$7.65$}       & & \footnotesize{$\sqrt{2}$ }  & \footnotesize{$2.55$}  &   \footnotesize{$7.65$}   \\
            \midrule
           \multirow{5}{*}{ Set5}          \\
           &  OneNet&     32.18  & 31.54 & 27.86 &        & 30.41& 29.52& 27.20          \\
           &  GS-PnP&     32.84 & 32.31 & \textbf{30.93} &        & 31.04& 29.98& \textbf{29.76}          \\
           &  RED&     32.63 & 31.76 & 29.83 &        & 31.25& 30.62 & 29.21          \\
        &  PnP-ADMM&     32.83  & 32.06 & 30.43 &        & 31.55& 30.88 & 29.30         \\

        &  \rev{PnP-ReG (Ours)}&     \textbf{33.40}  & \textbf{32.51} & 30.63 &        & \textbf{31.96} & \textbf{31.19} &  29.46        \\
           
            \midrule
          \multirow{5}{*}{ CBSD68}  \\        
          &  OneNet&     28.47  & 28.27 & 25.86 &        & 26.91& 26.97&  25.29          \\
            &  GS-PnP&     28.98 & \textbf{28.64} & \textbf{27.64 }&        & 27.5 & 27.33 & \textbf{26.57}          \\
            &  RED&     29.02 & 28.16 & 26.91 &        & 27.63& 27.24 & 26.20          \\
        &  PnP-ADMM&     29.21  & 28.54 & 27.20 &        & 27.91 & 27.40  & 26.25         \\

        &   \rev{PnP-ReG (Ours)}&     \textbf{29.38}  & 28.62 & 27.07 &        &\textbf{27.99} & \textbf{27.41}& 26.21        \\
        \midrule
         & &  \multicolumn{3}{c}{\includegraphics[width=0.08\linewidth]{kernel5.PNG} } & & \multicolumn{3}{c}{
   \includegraphics[width=0.08\linewidth]{kernel7.PNG} }  \\
            \cmidrule{3-5} \cmidrule{7-9} 
  &      $\sigma_n*255$         &  \footnotesize{$\sqrt{2}$}   & \footnotesize{$2.55$}  &   \footnotesize{$7.65$}       & & \footnotesize{$\sqrt{2}$ }  & \footnotesize{$2.55$}  &   \footnotesize{$7.65$}   \\
            \midrule
          \multirow{5}{*}{ Set5}         \\ 
          &  OneNet&     29.26 & 28.90 & 26.71 &        & 28.93& 28.18& 26.76          \\
         &  GS-PnP&     30.18 & 29.30 & \textbf{29.13} &        & 29.70& 29.07& \textbf{28.55}          \\
         &  RED&     30.53 & 29.98 & 28.67 &        & 29.91& 29.53 & 28.24          \\
        &  PnP-ADMM&     31.21  & 30.56 & 28.95 &        & 30.33 & 29.07 & 28.17         \\

       &   \rev{PnP-ReG (Ours)}&     \textbf{31.26}  & \textbf{30.57} & 28.84 &        & \textbf{30.67} & \textbf{29.95} &  28.31        \\
           
            \midrule
          \multirow{5}{*}{ CBSD68}    \\      
           &  OneNet&     26.52  & 26.57 & 25.04 &        & 25.88 & 25.94 &  24.77          \\
           &  GS-PnP&     27.02 & 27.04 & \textbf{26.32 }&        & 26.27 & 26.41 & \textbf{25.72}          \\
         &  RED&     27.36 & 26.99 & 26.00 &        & 26.61& 26.36 & 25.53          \\
       &  PnP-ADMM&     \textbf{27.83}  & \textbf{27.32} & 26.16 &        & 26.9 & 26.46  & 25.46         \\

       &   \rev{PnP-ReG (Ours)}&     27.71  & 27.16 & 25.96 &        &\textbf{26.93} & \textbf{26.47}& 25.47        \\
            \bottomrule[1.2pt]
        \end{tabu}
    }
    
    \label{tab:Deblurring PnP}
    \vspace{-8pt}
\end{table*}}

\begin{figure}
\centering
\setlength\tabcolsep{2pt} 

\begin{tabular}{c c c }
            Degraded Image   & One-Net  & \rev{GS-PnP} \\
            \includegraphics[width=0.23\linewidth]{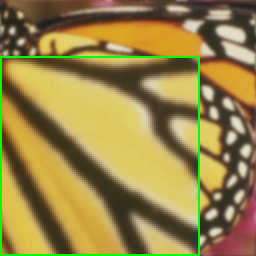} &
            \includegraphics[width=0.23\linewidth]{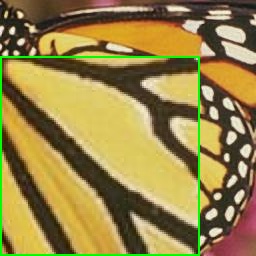} &
            \includegraphics[width=0.23\linewidth]{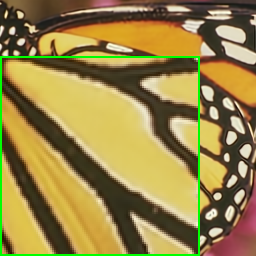}  \\
            &  32.50 dB  & \rev{34.33 dB} \\
        \end{tabular}

\centering
\begin{tabular}{c c c}
            RED   & PnP-ADMM  &  \rev{PnP-ReG} (Ours) \\
            \includegraphics[width=0.23\linewidth]{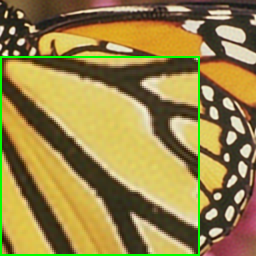} &
            \includegraphics[width=0.23\linewidth]{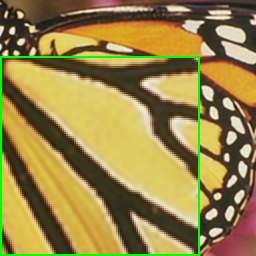} &
            \includegraphics[width=0.23\linewidth]{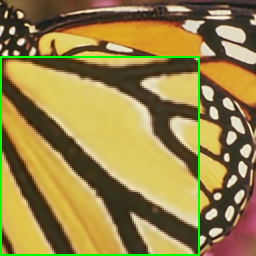}
             \\ 
              34.58 dB & 34.34 dB & 34.87 dB \\
        \end{tabular}

  \caption{\small{Visual comparison of deblurring results obtained with the projection operator (One-Net) \cite{rick2017one}, \rev{GS-PnP \cite{hurault2021gradient},}  RED \cite{romano2017little}, PnP-ADMM \rev{with the denoiser of} \cite{zhang2021plug} and our our PnP-ReG method. The blurred images have been generated using an isotropic Gaussian kernel of standard deviation $\sigma_b=1.6$ \rev{followed by adding Gaussian noise of standard deviation $\sigma_n = \sqrt{2}/255$}.
 }}
  \label{fig:PnP-DK2}
  \vspace{-10pt}

\end{figure}

\subsubsection{Pixel-wise inpainting}~\\ 
 Pixel-wise inpainting consists of restoring pixel-values in an image where a number of the pixels were randomly dropped. The degradation consists of multiplying the ground-truth image by a binary mask. We have tested our results with both 20\% and 10\% of known pixels rates. For the initialization image $\widehat x^0$, we have set the color of the unknown pixels to grey. 

\Cref{tab:Completion PnP} shows a numerical evaluation of our method for the application of pixel-wise inpainting in terms of PSNR [dB]. We observe significant performance gains of the PnP-ReG method compared to the other methods, especially in the most challenging case where the known pixel rate is only 10\%. Visual comparisons are given in \Cref{fig:PnP-Comp}.


{\renewcommand{\arraystretch}{0.9}

 \begin{table*}[h]
\caption{{\footnotesize{ Pixel-wise inpainting results (measured in PSNR [dB]) obtained with our PnP-ReG method; The corrupted images have been generated by keeping 20\% and 10\% of the known pixels. Comparison with the projection operator (One-Net) \cite{rick2017one}, \rev{ the Implicit Prior (IP) of \cite{kadkhodaie2020solving}, GS-PnP \cite{hurault2021gradient},} RED \cite{romano2017little} and PnP-ADMM \rev{with the denoiser of} \cite{zhang2021plug}.}}}
 \vspace{5pt}
\setlength\tabcolsep{5pt} 
    \centering
    { \normalsize
        \begin{tabu}{l l c c c c c}
            \toprule[1.2pt]
            & &  \multicolumn{2}{c}{Set5 } & & \multicolumn{2}{c}{CBSD68}  \\
            \cmidrule{3-4} \cmidrule{6-7} 
            &               & \footnotesize{(i) $10\%$}  &   \footnotesize{(ii) $20\%$}       &   & \footnotesize{(i) $10\%$} &   \footnotesize{(ii) $20\%$}   \\
            \midrule
                   
           &  OneNet&      17.20 & 24.06 &        &  14.72& 20.46          \\
          &  IP (avg)&      25.28 & 28.91 &  &24.12&  26.55          \\
          &  GS-PnP&    25.08 & 28.70 &        & 23.58& 25.91          \\
           &  RED&     22.75 & 27.17  &        &  22.24 & 23.22          \\
&  PnP-ADMM&     26.20  & 30.20  &        & 24.06& 26.75       \\

&   \rev{PnP-ReG (Ours)}&    \textbf{26.94} & \textbf{30.36} &        &  \textbf{24.43} &  \textbf{27.09}        \\
           
            \bottomrule[1.2pt]
        \end{tabu}
    }
    
    \label{tab:Completion PnP}
    \vspace{-8pt}
\end{table*}}

\vspace{5pt}

\begin{figure*}
\centering
\setlength\tabcolsep{2pt} 
\begin{tabu}{c c c c c c}
           Original image & \specialcell{Degraded\\image}    & One-Net 
           & \rev{IP (avg)} \\ 
           \includegraphics[width=0.22\linewidth]{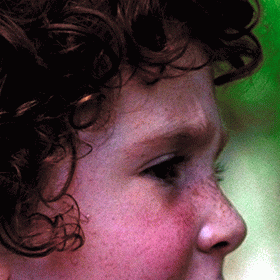} &
            \includegraphics[width=0.22\linewidth]{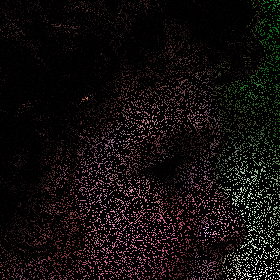} &

            \includegraphics[width=0.22\linewidth]{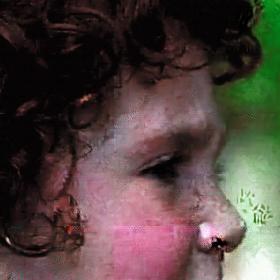} &

            \includegraphics[width=0.22\linewidth]{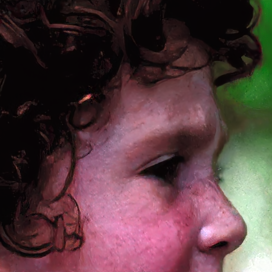} 
            \\
             &  & 25.06 dB & \rev{29.78 dB}\\
              \rev{GS-PnP} & RED & PnP-ADMM &  \rev{PnP-ReG} (Ours) \\

            \includegraphics[width=0.22\linewidth]{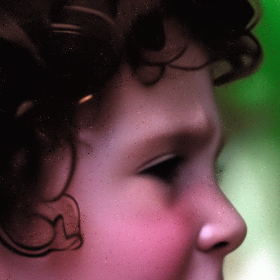} &
            \includegraphics[width=0.22\linewidth]{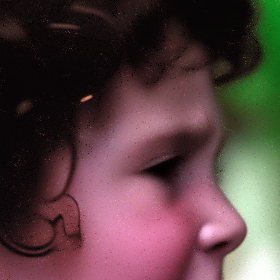} &
            \includegraphics[width=0.22\linewidth]{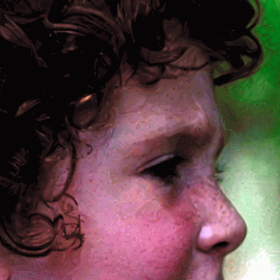} &
            \includegraphics[width=0.22\linewidth]{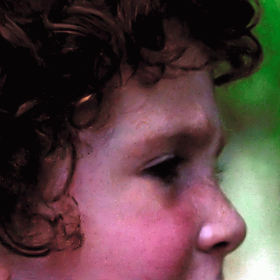} \\
            \rev{29.10 dB} & 28.11 dB & 29.60 dB & 29.82 dB\\

           Original image & \specialcell{Degraded\\image}    & One-Net 
           & \rev{IP (avg)} \\ 
           \includegraphics[width=0.22\linewidth]{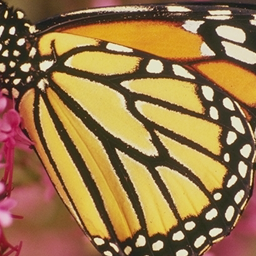} &
            \includegraphics[width=0.22\linewidth]{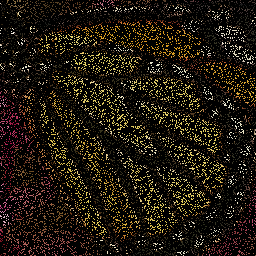}&

            \includegraphics[width=0.22\linewidth]{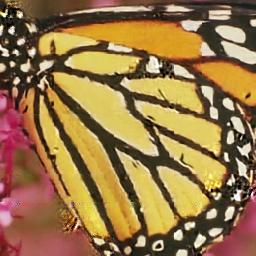}&

            \includegraphics[width=0.22\linewidth]{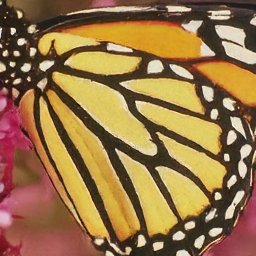} 
            \\
            &  & 20.65 dB & \rev{24.41 dB}\\
             \rev{GS-PnP} & RED & PnP-ADMM &  \rev{PnP-ReG} (Ours) \\

            \includegraphics[width=0.22\linewidth]{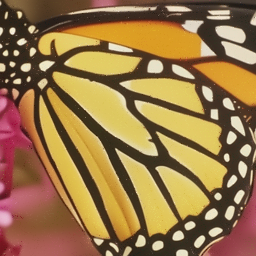}&
            \includegraphics[width=0.22\linewidth]{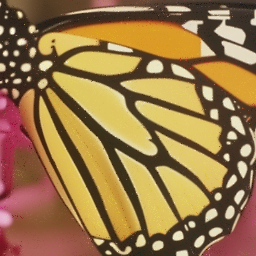}&
            \includegraphics[width=0.22\linewidth]{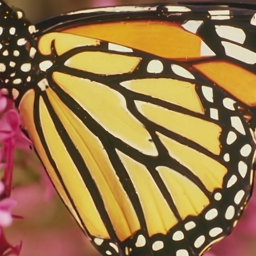} &
            \includegraphics[width=0.22\linewidth]{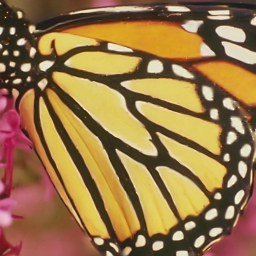} \\
          \rev{24.84 dB} & 23.51 dB & 24.99 dB & 25.51 dB\\
            
        \end{tabu}
  \caption{\small{Visual comparison of pixel-wise inpainting results with known pixel rate of $p=20\%$, obtained with the projection operator (One-Net) \cite{rick2017one}, \rev{ the Implicit Prior (IP) of \cite{kadkhodaie2020solving}, GS-PnP \cite{hurault2021gradient},} RED \cite{romano2017little}, PnP-ADMM \rev{with the denoiser of} \cite{zhang2021plug} and our PnP-ReG method.}
 }
  \label{fig:PnP-Comp}
  \vspace{-10pt}
\end{figure*}

\subsection{Unrolled gradient descent with \rev{$\mathcal{G}$}}
Aside from the Plug-and-Play gradient descent, our approach for training \rev{$\mathcal{G}$} can also serve as a pre-training strategy for unrolled gradient descent. In unrolled optimization methods, the regularization network is trained for each inverse problem such that applying a fixed number of iterations of the algorithm (e.g. \eqref{eq:GD1} for gradient descent) best approximates the ground truth image. We describe the unrolled training approach in \Cref{alg:Unrolled} for a gradient descent optimization. While this end-to-end training strategy loses the genericity of the Plug-and-Play approach, it typically improves the performances.

However, to facilitate the training, it is generally required to initialize the network weights with a generically pre-trained version. When unrolling proximal algorithms such as ADMM, a pre-trained deep denoiser can be used since it can be interpreted as the proximal operator of a generic regularization function. On the other hand, the gradient descent requires instead the gradient of a regularizer. Hence, a denoiser cannot be directly used as a pre-trained network. By transferring the image prior implicitly represented by the denoiser $\mathcal{D}$ to our ReG network \rev{$\mathcal{G}$}, our method thus provides a satisfying pre-trained network for unrolled gradient descent.

 \begin{table*}[h]
 \caption{ \footnotesize
        {Parameters used for unrolled gradient descent optimization for both the pre-trained and not pre-trained versions (i) Super-Resolution of factor 2 and 3 with input images corrupted using a bicubic filter  (ii) Deblurring images corrupted using an isotropic Gaussian kernel of standard deviation $\sigma_b$ of 1.6 and 2.0.  $\sigma_n$: Standard deviation of the White Gaussian noise added on the corrupted image, $\sigma$: weight of the regularization term, $\mu$: gradient step size , $N$: number of unrolled iterations. The training is performed over the DIV2K dataset \cite{agustsson2017ntire}.}}
 \vspace{8pt}
    \centering
    { \normalsize
        \begin{tabular}{r l r cc r cc}
            \toprule
            & & & \multicolumn{2}{c}{(i) Super-Resolution } & & \multicolumn{2}{c}{(ii) Deblurring}  \\
            \cmidrule{4-5} \cmidrule{7-8} 
            &               & & x2   & x3       & & $\sigma_b=1.6$   & $\sigma_b=2.0$       \\
            \midrule
            & $\sigma_n*255 $& &    0  & 0         & & $\sqrt{2}$ &  $\sqrt{2}$          \\
           & $\sigma*255$        & & 1.2       & 1.6         & & $\sqrt{2}$     & $\sqrt{2}$     \\
            & $\mu$          & & 0.008        & 0.1         & & 0.004       & 0.004    \\
            & $N$          & & 6       & 6         & & 6       & 6      \\

            \bottomrule
        \end{tabular}
    }
    
    \label{tab:parameters}
    \vspace{-8pt}
\end{table*}
 
\begin{algorithm}
\caption{Unrolled gradient descent with \rev{$\mathcal{G}$}}\label{alg:Unrolled}
\normalsize
\begin{algorithmic}[1]
\State initialize $ \rev{\mathcal{G}} , \sigma_n, \mu,  A, B$ 
\For{each epoch}      
    \For{each batch}  
    \State $\eta \gets \mathcal{N}(0,\sigma_n)$ 
       \State  $x_{gt} \gets$ ground-truth batch 
       \State $y \gets Ax_{gt} + \eta$ 
       \State $x_0 \gets y$
       \For{$k \gets 0$ to $N-1$}   
            \State $x_{k+1} \gets x_k - \mu [ A^T ( Ax_k - y) + \sigma^2 \rev{\mathcal{G}} (x_k)] $
        \EndFor
        \State loss $ \gets \left\| x_N-x_{gt} \right\|_2 ^2$
        \State Update weights of \rev{$\mathcal{G}$} with back-propagation 
   \EndFor
\EndFor
\end{algorithmic}
\end{algorithm}
We have tested this approach for super-resolution of factor 2 and 3 (without noise), as well as for deblurring using 2 isotropic Gaussian kernels (with standard deviations of 1.6 and 2.0) \rev{and additive Gaussian noise of standard deviation $\sqrt{2}$/255}. We have compared our results with those obtained when using a network learned end-to-end in an unrolled environment without the pre-training (i.e. random weight initialization).
 
Both versions (pre-trained and not pre-trained) have been unrolled in the same conditions.  \Cref{tab:parameters} gives the training parameters for each of the different tasks. For all 4 cases, we have trained the networks using the DIV2K dataset \cite{agustsson2017ntire} of 800 images, over 600 epochs by randomly taking 48x48 patches from the dataset.
In addition, we include a comparison with the Total Deep Variation (TDV) method \cite{Kobler2020TDV} which also performs unrolled optimization where the network represents the gradient of the regularization function. In \cite{Kobler2020TDV}, the network is not pre-trained. Instead, the training starts with a small number of unrolled iterations $N=2$, and $N$ is incremented every 700 epochs. Note that the authors originally trained the TDV network for $N=10$ iterations of an unrolled proximal gradient descent algorithm. However, for fair comparisons with our approach, we re-trained it with $N=6$ iterations of simple unrolled gradient descent.

\begin{figure*}
\centering
\setlength\tabcolsep{2pt} 
\begin{tabular}{ccccc}
            \multirow{2}{*} {Ground Truth } &  \multirow{2}{*} {\specialcell{Bicubic\\interpolation}}   & Unrolled GD  & Unrolled GD & \multirow{2}{*} {TDV }\\
             &  & (pre-trained) & (not pre-trained) \\

            \includegraphics[width=0.19\linewidth]{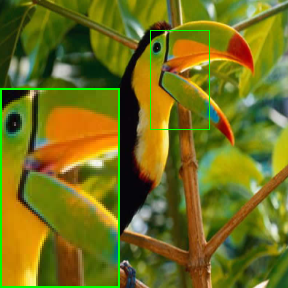} &
            \includegraphics[width=0.19\linewidth]{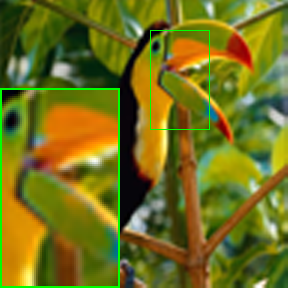} & 
            \includegraphics[width=0.19\linewidth]{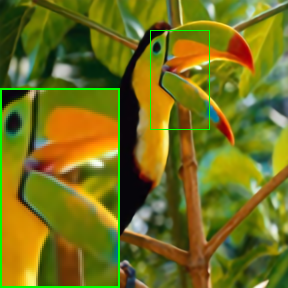} &
            \includegraphics[width=0.19\linewidth]{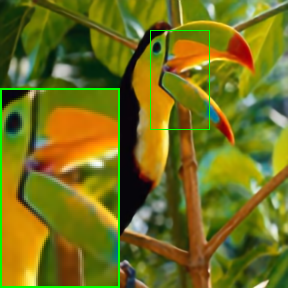}
            &
            \includegraphics[width=0.19\linewidth]{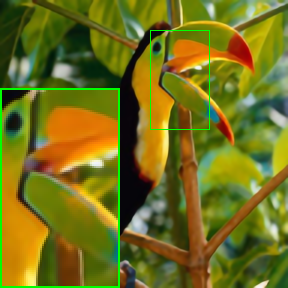}\\

             & 30.73 dB & 35.10 dB & 34.56 dB & 34.03 dB\\
            
        \end{tabular}

  \caption{\small
        {Visual comparison of super-resolution results between unrolled gradient descent with and without pre-training. Low resolution images have been generated with a bicubic kernel followed by a downsampling by a factor of 3.}}
  \label{fig:Unrolled-SR3}
  \vspace{-10pt}
\end{figure*}

\begin{figure*}
\centering
\setlength\tabcolsep{2pt} 
\begin{tabular}{ccccc}

            \multirow{2}{*} {Ground Truth } & \multirow{2}{*} {Blurry image }  & Unrolled GD  & Unrolled GD &  \multirow{2}{*} {TDV }\\
             &  & (pre-trained) & (not pre-trained) &  \\
            \includegraphics[width=0.19\linewidth]{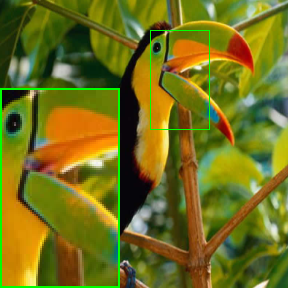} &
            \includegraphics[width=0.19\linewidth]{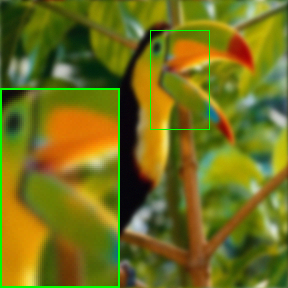} &
            \includegraphics[width=0.19\linewidth]{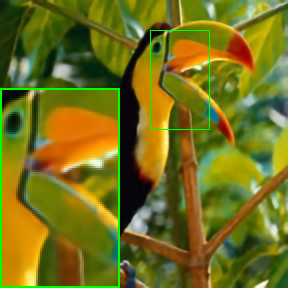} &
            \includegraphics[width=0.19\linewidth]{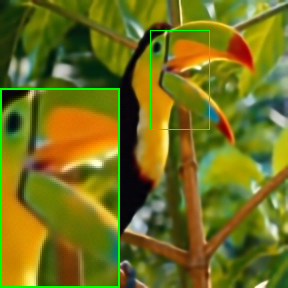} &
            \includegraphics[width=0.19\linewidth]{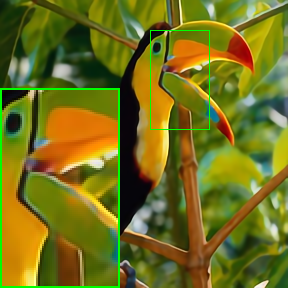}\\
            & & 33.83 dB & 32.07 dB & 32.48 dB \\
           \includegraphics[width=0.19\linewidth]{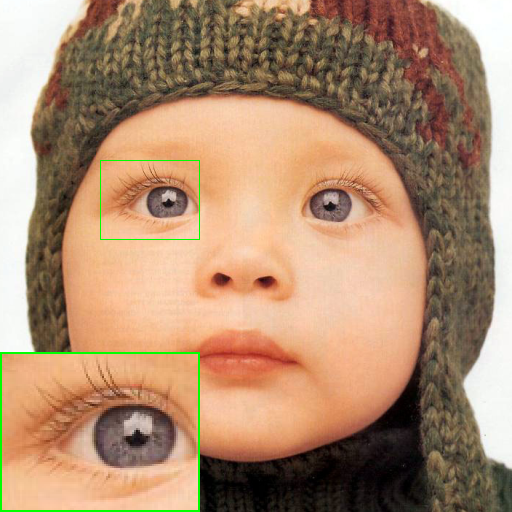} &
            \includegraphics[width=0.19\linewidth]{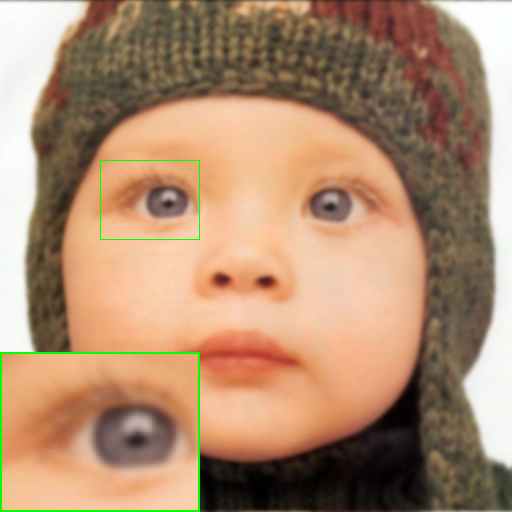} &
            \includegraphics[width=0.19\linewidth]{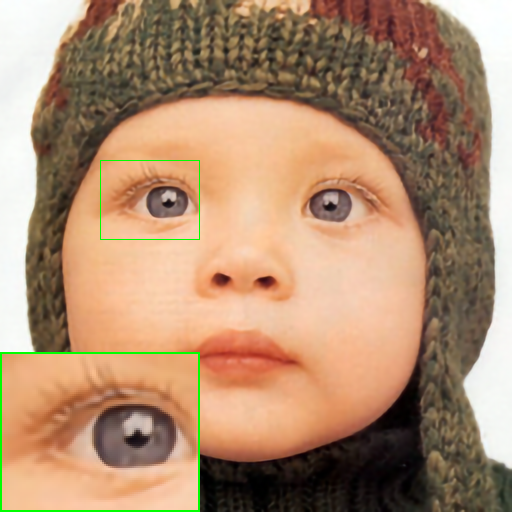} &
            \includegraphics[width=0.19\linewidth]{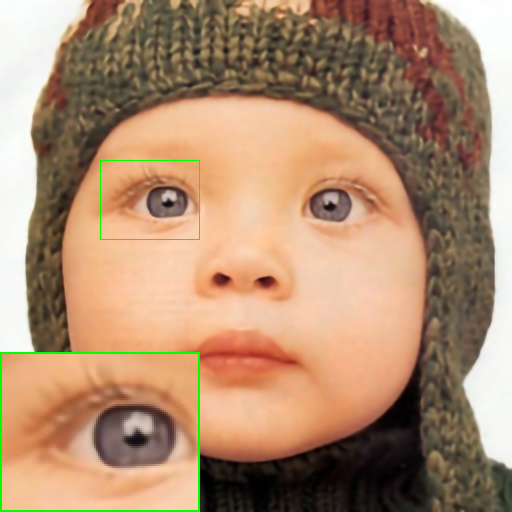} &
            \includegraphics[width=0.19\linewidth]{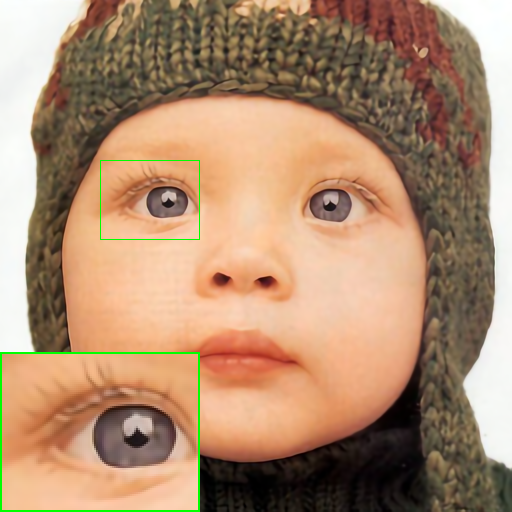}\\
            & & 34.13 dB & 33.31 dB & 33.89 dB \\
            
        \end{tabular}

  \caption{\small{Visual comparison of deblurring results between unrolled gradient descent with and without pre-training. The blurred images have been generated using an isotropic Gaussian kernel of standard deviation 2.0.}}
  \label{fig:Unrolled-Deb}
  \vspace{-10pt}
\end{figure*}

\Cref{tab:SR Unrolled,tab:Deblurring Unrolled} show the PSNR results [dB] for both networks (pre-trained and not pre-trained) as well as for the TDV, for super-resolution and deblurring respectively, using Set5 \cite{Set5}, Set14 \cite{Set14} and BSDS100\cite{CBSD68}. As expected, using \rev{$\mathcal{G}$} for weight initialization improves the results of the unrolled gradient descent for the 4 tested cases (with up to 0.9 dB).
Some visual comparisons of super-resolution results with magnifying factor of 3, and of deblurring images corrupted by a Gaussian kernel of standard deviation 2.0 are respectively shown in  \Cref{fig:Unrolled-SR3,fig:Unrolled-Deb}. The visual results confirm that better reconstruction of the details is obtained when our pre-training is used. While the TDV results display even sharper edges, the PSNR remains lower because of exaggerated sharpness in comparison to the ground truth.

\begin{table}[h]
\caption{\footnotesize{{ Super-resolution results (measured in PSNR [dB]) obtained with unrolled gradient descent (the input images have been corrupted using a bicubic kernel and a downsampling factor of 2 and 3). The evaluation has been performed using the Set5, Set14 and BSDS100 datasets. The restoration has been performed using unrolled gradient descent initialized with our pre-trained network \rev{$\mathcal{G}$} and without weight initialization, as well as TDV \cite{Kobler2020TDV}}}}
\setlength\tabcolsep{4.5pt}
\vspace{5pt}
    \centering
    {\normalsize
        \begin{tabular}{r l cc cc cc}
            \toprule
           & && \specialcell{pre-trained\\(ours)} & \specialcell{not\\pre-trained} & TDV \\ 
            \midrule
            \multirow{2}{*} {\footnotesize  (i) x2 } 
&Set5  && 35.61& 35.42& 34.57 \\ 
&Set14 && 31.18& 30.93& 30.31 \\ 
&BSDS100 && 30.77 & 30.68& 30.23  \\ 
            \midrule
            \multirow{2}{*} {\footnotesize  (i) x3 } 
&Set5 & & 32.10& 31.88& 31.47  \\ 
&Set14 && 28.00& 27.79& 27.40  \\ 
&BSDS100 && 27.68 & 27.63& 27.34  \\ 
            \bottomrule
        \end{tabular}
    }
    
    \label{tab:SR Unrolled}
    \vspace{-8pt}
\end{table}
\begin{table}[h]
\caption{\footnotesize{{ Deblurring results (measured in PSNR [dB]) obtained with unrolled gradient descent (the input images have been corrupted using isotropic Gaussian kernels with standard deviation 1.6 and 2.0 \rev{and Gaussian noise of standard deviation $\sqrt{2}/255$}). The evaluation has been performed using the Set5, Set14 and BSDS100 datasets. The restoration has been performed using unrolled gradient descent initialized with our pre-trained network \rev{$\mathcal{G}$} and without weight initialization, as well as \cite{Kobler2020TDV}.}}}
\setlength\tabcolsep{4.5pt}
\vspace{5pt}
    \centering
    {\normalsize
        \begin{tabular}{r l cc cc cc}
            \toprule
           & && \specialcell{pre-trained\\(ours)} & \specialcell{not\\pre-trained} & TDV \\ 

            \midrule
            \multirow{3}{*} {\footnotesize (i) $\sigma_0=1.6$ } 
&Set5 & & 33.51& 32.57& 32.32  \\ 
&Set14 && 30.14& 29.33& 29.17  \\ 
&BSDS100 && 29.80 & 29.12& 29.01  \\ 
            \midrule
            \multirow{3}{*} {\footnotesize (ii) $\sigma_0=2$ } 
&Set5 && 31.63 & 30.94 & 30.52  \\ 
&Set14 && 28.29& 27.70 & 27.46  \\ 
&BSDS100 && 28.08 & 27.52 & 27.51  \\ 
            \bottomrule
        \end{tabular}
    }
    
    \label{tab:Deblurring Unrolled}
    \vspace{-8pt}
\end{table}

\subsection{Analysis of the joint training}
\label{ssec:joint_train}
In this section, we experimentally verify the advantages of updating the denoising network within the training of our ReG network \rev{$\mathcal{G}$}, compared to letting the denoiser fixed. In the latter case, $\delta$ is always set to $0$ (i.e. $\mathcal{L} = \lambda \mathcal{L}_{\rev{\mathcal{G}}}$).

\begin{figure}[h]
\centering
\begin{tabular}{cc}

   \includegraphics[width=0.42\linewidth]{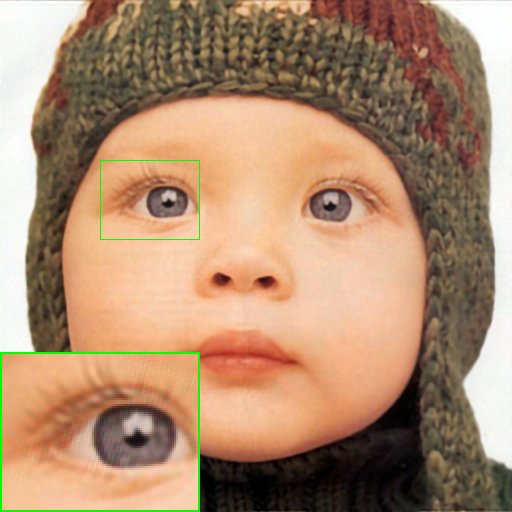} &
    \includegraphics[width=0.42\linewidth]{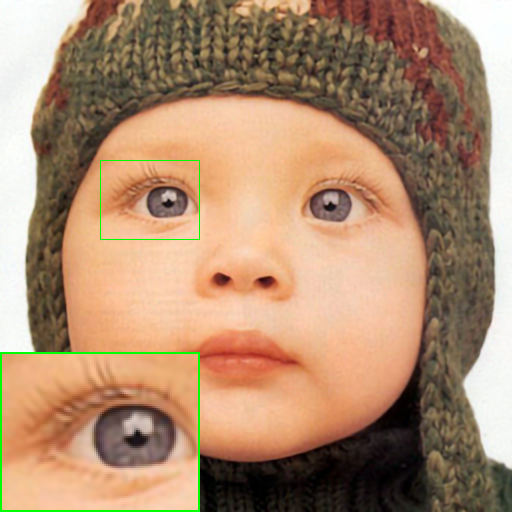}\\
     (a) 33.72 dB & (b) 35.46 dB \\

    \multicolumn{2}{c}{\includegraphics[width=0.5\linewidth]{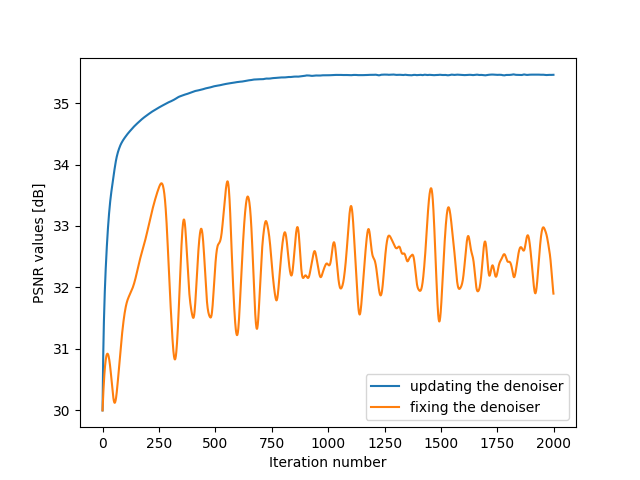} }\\
       \multicolumn{2}{c}{ (c) }\\
\end{tabular}
  \caption{\small{Comparison of the performances of PnP-ReG, when the ReG network is trained (a) with a fixed denoiser and (b) while jointly updating the denoiser. (c) PSNR values over the gradient descent iterations. The results are shown for the deblurring problem (the blurred images have been generated using isotropic Gaussian kernels of standard deviation 1.6 followed by adding Gaussian noise of standard deviation $\sqrt{2}/255$).}}
  \label{fig:den-gd}
\end{figure}

First, we compare the performance of our regularizing gradient network trained in both scenarios. An example of deblurring results with the Plug-and-Play gradient descent is shown in \Cref{fig:den-gd}. It is clear that leaving the denoiser fixed to its pre-trained state degrades the performance of \rev{$\mathcal{G}$}: the reconstructed image in \Cref{fig:den-gd} (a) remains more blurry than in \Cref{fig:den-gd} (b) and it also presents colored fringes artifacts.
On the other hand, when the denoiser is updated when training \rev{$\mathcal{G}$}, the convergence of the Plug-and-Play  gradient descent is significantly improved as well as the visual result, as shown in \Cref{fig:den-gd} (b,c).




A possible explanation of the worse results when fixing the denoiser in our training is that the pre-training of $\mathcal{D}_\sigma$ does not guarantee that there exists a differentiable regularizer $\phi$ for which $\mathrm{prox}_{\sigma^2\phi} = \mathcal{D}_\sigma$ for every value of $\sigma$. In other words, the assumption that $\mathcal{D}_\sigma$ is a MAP Gaussian denoiser for a differentiable prior may not be satisfied.
However, by jointly updating the denoiser with the ReG network, the modified denoiser better represents such a MAP Gaussian denoiser for the corresponding regularizer.

\begin{figure*}[h]
\centering
\begin{tabular}{cc}

   \includegraphics[width=0.45\linewidth]{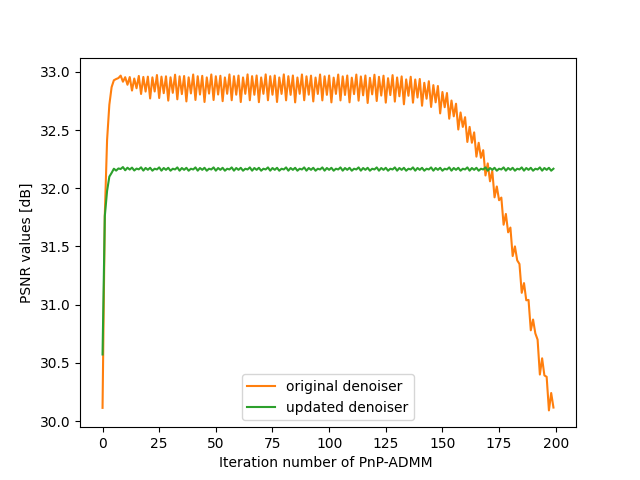} &
    
   \includegraphics[width=0.45\linewidth]{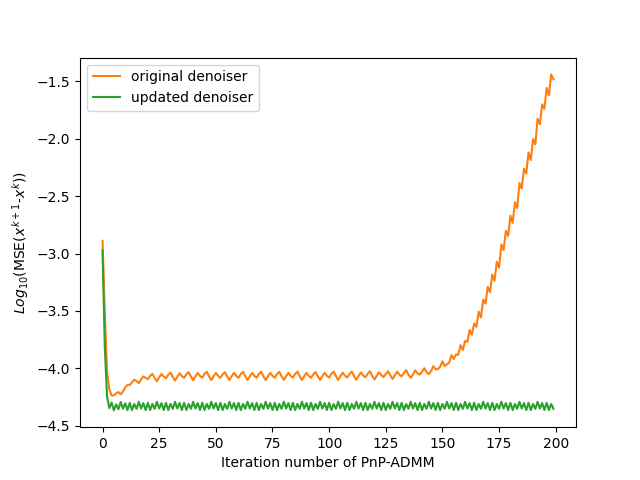} \\
(a) & (b) \\
\end{tabular}
  \caption{\small{Comparison of the performances of the original and the updated denoisers in PnP-ADMM for deblurring, using the Set5 dataset (The blurred
images have been generated using an isotropic Gaussian kernel of standard deviation 1.6 followed by adding Gaussian noise of standard deviation $\sqrt{2}/255$): (a) Average PSNR over the ADMM iterations (b) MSE of the difference between two consecutive iterations (in log scale).}}
  \label{fig:den-admm}
\end{figure*}

In a second experiment, we compare the performances of the updated denoiser and the original DRUNet when used in the Plug-and-Play ADMM algorithm.
\Cref{fig:den-admm} shows for each ADMM iteration the PSNR of the reconstructed images and the MSE of the difference between two consecutive iterations for both versions of the denoiser.
We can observe that although the original DRUNet can obtain better PSNR performances when stopping the ADMM after a few iterations (see subfigure (a)), the algorithm does not converge, and may even strongly diverge after a sufficiently large number of iterations. On the other hand our modified denoiser allows for a better convergence of the Plug-and-Play ADMM.


\section{Discussion}
\begin{figure*}[h]
\centering

   \includegraphics[width=0.65\linewidth]{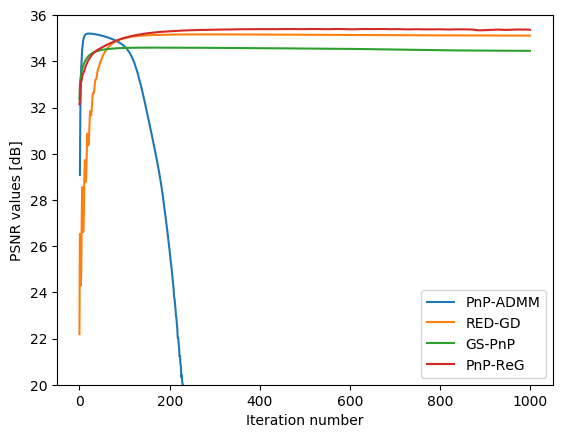} 
    
  \caption{\small{\rev{Comparison of the convergence of PnP-ADMM with the denoiser of \cite{zhang2021plug}, RED \cite{romano2017little}, GS-PnP \cite{hurault2021gradient} and PnP-ReG for Super-Resolution. The input low resolution images have been generated using bicubic downsampling with a factor 2. The results are averaged over the images of the Set5 dataset. For PnP-ADMM, we have used the setting with variable parameter $s^k$ until the $25^{th}$ iteration for which the best results are obtained (see Table \ref{tab:Parameters PnPADMM}), and let $s^k$ fixed afterwards.}}}
  \label{fig:convergence-SR}
\end{figure*}

\begin{figure*}[h]
\centering
   \includegraphics[width=0.65\linewidth]{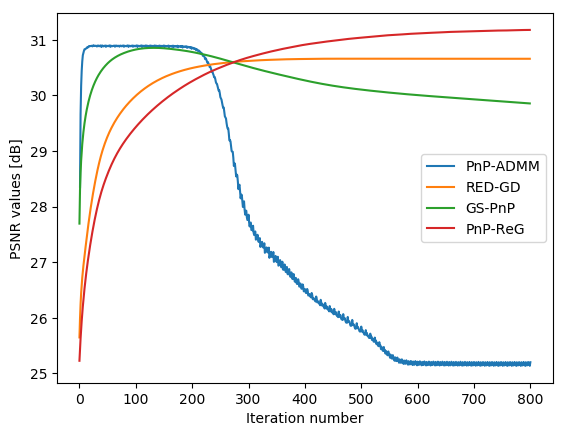} 

  \caption{\small{\rev{Comparison of the convergence of PnP-ADMM with the denoiser of \cite{zhang2021plug}, RED \cite{romano2017little}, GS-PnP \cite{hurault2021gradient} and PnP-ReG for Deblurring. The degraded images have been generated using an isotropic Gaussian kernel of standard deviation 2.0 followed by adding Gaussian noise of standard deviation $\sigma_n = 0.01$. The results are averaged over the images of the Set5 dataset. For PnP-ADMM, we have used the setting with fixed parameter $s^k = 30/255$ for which the best results are obtained (see Table \ref{tab:Parameters PnPADMM}).}}}
  \label{fig:convergence-deb}
\end{figure*}
\rev{In this section, we discuss the advantages and the limitations of our PnP-ReG method.
First, we point out that the proposed approach only requires the tuning of the gradient step $\mu$ when the noise level is known, while the reference PnP methods heavily rely on the tuning of several hyper-parameters, sometimes even including the number of iterations. This makes our method easier to use since parameter tuning can take a lot of effort. 

Furthermore, the comparison of the convergence of the different PnP algorithms in \Cref{fig:convergence-SR,fig:convergence-deb} show that although our method requires more iterations than PnP-ADMM and GS-PnP to reach its highest PSNR result, it converges to a fixed point that provides the best quality. On the other hand, PnP-ADMM reaches its highest PSNR after a few iterations, but then diverges, which requires a careful tuning of the number of iterations for each application in practice.
The convergence of our PnP-ReG method is more comparable to the RED-GD method (although significant PSNR gains are obtained with our method). This may be explained by the fact that both RED-GD and PnP-ReG are based on gradient descent while PnP-ADMM and GS-PnP use proximal algorithms.

Aside from the convergence analysis, one may notice that, even though we have designed our loss from \eqref{eq:m6} so that $\mathcal{G}$ models $\nabla\phi$, we did not enforce the network to have a symmetric Jacobian matrix. As a result, our training framework does not guarantee that $\mathcal{G}$ can be interpreted mathematically as a conservative vector field, and thus, as the gradient of a scalar potential.
However, using a non-conservative vector field for the gradient update step may not necessarily degrade the gradient descent performance. For example, advanced gradient-based algorithms typically alter the gradient (e.g. with momentum), in a way that loses the conservativeness property of the original gradient, while improving the algorithm's performances (more robust to local minima, faster convergence...).
Note also that it would be possible to enforce the Jacobian symmetry property in our method by using a network directly modeling $\phi$ and by explicitly computing its gradient, as done in \cite{cohen2021it,hurault2021gradient}. However, the explicit gradient computation has the same complexity as the network $\phi$, hence doubling the computation cost. A possible direction for future work would be to study whether such a constraint could improve the Plug-and-Play gradient descent without sacrificing the computational complexity.
}

\section{Conclusion}
In this paper, we have proposed a novel framework for solving linear inverse problems. Our approach makes it possible to solve Plug-and-Play algorithms using gradient descent, where the gradient of the regularizer is required rather than its proximal operator. We have proved that it is mathematically possible to train a network that models the gradient of a regularizer, jointly with a denoising neural network. 

The results have demonstrated that the joint training of our network with a DRUNet denoiser has several advantages. First, our network can be used to regularize the gradients in a Plug-and-Play gradient descent algorithm and can outperform other generic approaches in different inverse problems such as super-resolution, deblurring and  pixel-wise inpainting. Second, our network can also serve as a pre-training strategy for unrolled gradient descent and yield a significant improvement. Lastly, the joint training of the denoiser with the regularizing gradient network makes the former match better the definition of a proximal operator compared to the original pre-trained DRUNet.


\appendix
\section{Plug-and-Play ADMM: Formulation and parameter setting} \label{sec:pnpadmm}
\rev{
While the Plug-and-Play ADMM can provide high quality image reconstruction, the parameter setting is not trivial and strongly influences the results. We detail in this section how we have parameterized the Plug-and-Play ADMM method for our comparisons.
First, let us describe the ADMM equations for solving linear inverse problems as defined in \eqref{eq:MAP2}. The formulation is derived by first decoupling the data term and the prior term by adding an auxiliary variable $\vect{z}$, yielding a constrained optimization problem which is equivalent to \eqref{eq:MAP2}:
\begin{equation}\label{eq:prop1}
\begin{aligned}
\vect{\hat{x}} = &\argmin_{\vect{x},\vect{z}} & & \frac{1}{2} \norm{\matr{A}\vect{x}-\vect{y}}_2^2 + \sigma^2 \mathcal{\phi}(\vect{z}).\\
&\text{subject to} & & \vect{x}=\vect{z},
\end{aligned}
\end{equation}


Note that $\sigma$ is a parameter of the problem to be solved and should not depend on the algorithm. As discussed in \Cref{ssec:ParamSettings}, $\sigma$ must be equal to the true noise level $\sigma_n$ added in the degradation. However, in the noiseless scenario, using $\sigma=\sigma_n=0$ removes the regularization term. Hence similarly to our method, we choose a very small non-zero value in this case. For our experiments, we have used $\sigma=\max(\sigma_n, 0.001/255)$.

Each ADMM iteration then consists in an alternate minimization over $\vect{x}$ and $\vect{z}$ with the introduction of an additional variable $\vect{l}$ called dual variable as well as a penalty parameter $\rho$, as follows:

\begin{align}
\label{eq:data_subpb}
\vect{x^{k+1}} &= \argmin_{\vect{x}} \norm{\matr{A}\vect{x}-\vect{y}}_2^2 + \rho^k\norm{\vect{x}-\left(\vect{z^k}-\frac{\vect{l^k}}{\rho^k}\right)}_2^2,\\
\label{eq:prior_subpb}
\vect{z^{k+1}} &= \argmin_{\vect{z}} \frac{1}{2}\norm{\vect{z}-\left(\vect{x^{k+1}}+\frac{\vect{l^k}}{\rho}\right)}_2^2 + \frac{\sigma^2}{\rho}\phi(\vect{z}) = \prox_{\frac{\sigma^2}{\rho}\phi}(\vect{x^{k+1}}+\vect{l^k}/\rho),\\
\label{eq:dual_update}
\vect{l^{k+1}} &= \vect{l^k} + \rho(\vect{x^{k+1}}-\vect{z^{k+1}}),
\end{align}
where the dual variable is typically zero-initialized.

As discussed earlier in \Cref{sec:notations},
proximal operators can be interepreted as MAP Gaussian denoisers. Formally, a MAP Gaussian denoiser for a given noise standard deviation $s$ and some underlying regularizer $\phi$ can be written $\mathcal{D}_{s}=\prox_{s^2\phi}$. The Plug-and-Play ADMM thus replaces the $z$-update in \eqref{eq:prior_subpb} by the following denoising step: 
\begin{equation}
    \vect{z^{k+1}} = \mathcal{D}_{s}(\vect{x^{k+1}}+\vect{l^k}/\rho),
\end{equation}
where $s=\sqrt{\sigma^2/\rho}=\sigma/\sqrt{\rho}$. Note that the PnP-ADMM may thus parameterize the denoiser with a different standard deviation $s$ than the noise level $\sigma$ by setting the penalty parameter $\rho\neq1$. In practice, the results of PnP-ADMM strongly depend on the setting of $\rho$. A common strategy in ADMM is to increase the value of $\rho$ at each iteration with an update of the form $\rho^{k+1}=\rho^k \cdot \alpha$ for some fixed parameter $\alpha\geq1$. Hence, this requires setting two parameters $\rho_0$ and $\alpha$.
A more intuitive parameterization typically used in the Plug-and-Play context (e.g. \cite{zhang2021plug, lependu2021preco}) considers instead  the denoising standard deviations $s^0$ and $s^N$ respectively at the first and last iterations. Knowing that $s=\sigma/\sqrt{\rho^k}$ at each iteration, $\rho^0$ and $\alpha$ can be set accordingly as:
\begin{equation}
\label{eq:params-setting}
\rho^0=\left(\frac{\sigma}{s^0}\right)^2 \quad\text{and}\quad \alpha=\left(\frac{s^0}{s^N}\right)^{2/N}
\end{equation}





Therefore, the main parameters that we need to set in PnP-ADMM are the number of iterations $N$ and the initial and final noise standard deviation of the denoiser (i.e. $s^0$ and $s^N$). For our experiments, we have tested two configurations: the first configuration is similar to \cite{zhang2021plug} and \cite{lependu2021preco} with $s$ varying between a sufficiently high value $s^0$ and the true noise standard deviation $s^N=\sigma_n$ (or a small value when $\sigma_n=0$). In the second configuration, $s$ is kept constant (i.e. $s^0=s^N$). For both configurations, we have tuned the parameters to obtain the best results on the Set5 dataset, and we kept the best configuration. The parameters are given in Table \ref{tab:Parameters PnPADMM} in Appendix \ref{sec:parameters}.


\section{Parameter setting and implementation details for the other methods} \label{sec:parameters}

{\renewcommand{\arraystretch}{0.9}
\begin{table}[h]
\caption{{\footnotesize{Parameters used for the projection operator (One-Net) \cite{rick2017one}. $\sigma_n$: standard deviation of the AWGN added on the degraded image, $\rho$: penalty parameter of the ADMM, $n$: number of iterations}}}
\setlength\tabcolsep{4pt} 
\vspace{5pt}
    \centering
    {\normalsize
        \begin{tabular}{l l c |c c c}
            \toprule
            && $\sigma_n*255$ & \hspace*{0.5cm}&$\rho$ & $n$   \\ 
            \midrule
            \multirow{6}{*}{Super-Resolution x2} 
            & \multirow{3}{*}{Bicubic} 
             & 0 & &0.05 & 1 \\ 
              && $\sqrt{2}$ & &0.06 & 12 \\ 
             && 2.55 & &0.06 & 12 \\
             \cmidrule{2-6}
    &\multirow{3}{*}{Gaussian}                  
            & 0 & &0.02 & 1 \\ 
              && $\sqrt{2}$ & &0.06 & 10 \\ 
              && 2.55 & &0.06 & 10 \\
    \midrule
     \multirow{6}{*}{Super-Resolution x3} & 
                \multirow{3}{*}{Bicubic} 
            & 0 & &0.01 & 5 \\ 
              && $\sqrt{2}$ && 0.06 & 10 \\ 
              && 2.55 && 0.05 & 12 \\
             \cmidrule{2-6}
    &\multirow{3}{*}{Gaussian}     
    & 0 & &0.01 &5 \\ 
              && $\sqrt{2}$ & &0.07 & 10 \\ 
              && 2.55 & &0.07 & 10 \\
    \midrule
         \multirow{3}{*}{Deblurring} & 
                \includegraphics[width=0.03 \linewidth]{kernel2.PNG}  \includegraphics[width=0.03\linewidth]{kernel3.PNG}
                \includegraphics[width=0.03 \linewidth]{kernel5.PNG}  \includegraphics[width=0.03\linewidth]{kernel7.PNG}
             & $\sqrt{2}$ & &0.01 & 15 \\ 
             &\includegraphics[width=0.03 \linewidth]{kernel2.PNG}  \includegraphics[width=0.03\linewidth]{kernel3.PNG}
                \includegraphics[width=0.03 \linewidth]{kernel5.PNG}  \includegraphics[width=0.03\linewidth]{kernel7.PNG} & 2.55 && 0.02 & 15 \\ 
             & \includegraphics[width=0.03 \linewidth]{kernel2.PNG}  \includegraphics[width=0.03\linewidth]{kernel3.PNG}
                \includegraphics[width=0.03 \linewidth]{kernel5.PNG}  \includegraphics[width=0.03\linewidth]{kernel7.PNG}& 7.65 && 0.02 & 5 \\
             
    \midrule
    \multirow{2}{*}{Pixel-wise inpainting } & $0.1$ & 0 &&0.004 & 300 \\ 
    &$0.2$       & 0&& 0.01 & 250 \\
            \bottomrule
        \end{tabular}
    }
    \label{tab:Parameters OneNet}
    \vspace{-8pt}
\end{table}}
\vspace{5pt}

{\renewcommand{\arraystretch}{0.9}
\begin{table}[h]
\caption{{\footnotesize{Parameters used for the RED \cite{romano2017little} in a Gradient Descent framework using DRUNet. $\sigma_n$: standard deviation of the AWGN added on the degraded image, $w$: weight of the regularization, $\sigma_f$: noise standard deviation of the denoiser, $\mu$: gradient step size. }}}
\setlength\tabcolsep{4pt} 
\vspace{5pt}
    \centering
    {\normalsize
        \begin{tabular}{l c c | c c c c}
            \toprule
           & & $\sigma_n*255$ & \hspace*{0.5cm} & $w$ & $\sigma_f*255$ & $\mu$   \\ 
            \midrule
            \multirow{3}{*}{\specialcell{Super-Resolution x2\\ (Bicubic and Gaussian)}} 
              && 0 & &0.005 & 7 & 0.08 \\ 
              && $\sqrt{2}$ && 0.03 & 7 & 0.08\\ 
              && 2.55 && 0.07 & 7 & 0.08\\
             
    \midrule
     \multirow{3}{*}{\specialcell{Super-Resolution x3\\ (Bicubic and Gaussian)}}  & 
            & 0 & &0.005 & 10 & 0.08\\ 
              && $\sqrt{2}$ && 0.01 & 10 & 0.08\\ 
              && 2.55 && 0.03 & 10 & 0.08\\
            
    \midrule
         \multirow{6}{*}{Deblurring} & 
         \multirow{3}{*}{\includegraphics[width=0.05 \linewidth]{kernel2.PNG}}

             & $\sqrt{2}$ &&0.01 & 10 & 0.01\\ 
            
             && 2.55 && 0.03 & 16 & 0.01\\ 
            && 7.65 && 0.03 & 16 & 0.01 \\
             \cmidrule{2-7}
            & \multirow{3}{*}{\includegraphics[width=0.05 \linewidth]{kernel3.PNG}
            \includegraphics[width=0.05 \linewidth]{kernel5.PNG}
            \includegraphics[width=0.05 \linewidth]{kernel7.PNG}}

             & $\sqrt{2}$ && 0.01 & 16 & 0.01\\ 
            
             && 2.55 && 0.01 & 16 & 0.01\\ 
            && 7.65 && 0.03 & 16 & 0.01 \\

    \midrule
    \multirow{2}{*}{Pixel-wise inpainting } & $0.1$ & 0& &0.001 & 27 & 0.006\\ 
    & $0.2$       & 0 && 0.001 & 24 & 0.008 \\
            \bottomrule
        \end{tabular}
    }
    \label{tab:Parameters RED}
    \vspace{-8pt}
\end{table}}
\vspace{5pt}

{\renewcommand{\arraystretch}{0.9}
\begin{table}[h]
\caption{{\footnotesize{Parameters used for the PnP-ADMM with the denoiser of \cite{zhang2021plug}. $\sigma_n$: standard deviation of the AWGN added on the degraded image, $s^0$ noise standard deviation of the denoiser at the first iteration, $s^N$: noise standard deviation of the denoiser at the last iteration, $N$: number of iterations.}}}
\setlength\tabcolsep{4pt} 
\vspace{5pt}
    \centering
    {\normalsize
        \begin{tabular}{l l c | c c c c}
            \toprule
            && $\sigma_n*255$ & \hspace*{0.5cm} &$s^0*255$ & $s^N*255$ & $N$  \\ 
            \midrule
            \multirow{6}{*}{Super-Resolution x2} 
            & \multirow{3}{*}{Bicubic} 
            & 0 && 50 & 0.1 & 25\\ 
              && $\sqrt{2}$ && 20 & 20 & 30 \\ 
              && 2.55 && 30 & 30 & 30 \\
             \cmidrule{2-7}
    &\multirow{3}{*}{Gaussian}                  
             & 0 && 50 & 0.1 & 25\\ 
              && $\sqrt{2}$ && 30 & 30 & 30 \\ 
              && 2.55 && 45 & 45 & 30 \\
    \midrule
     \multirow{6}{*}{Super-Resolution x3} & 
                \multirow{3}{*}{Bicubic} 
             & 0 && 50 & 0.1 & 40\\ 
              && $\sqrt{2}$ && 60 & 60 & 30 \\ 
              && 2.55 && 100 & 100 & 30 \\
             \cmidrule{2-7}
    &\multirow{3}{*}{Gaussian}                  
             & 0 && 50 & 0.1 & 25\\ 
              && $\sqrt{2}$ && 30 & 30 & 30 \\ 
              && 2.55 && 45 & 45 & 30 \\
    \midrule
         \multirow{3}{*}{Deblurring} & 
                               \multirow{3}{*}{\includegraphics[width=0.05 \linewidth]{kernel2.PNG}  \includegraphics[width=0.05\linewidth]{kernel3.PNG}} 
            
              & $\sqrt{2}$ && 25 & 25 & 20 \\ 
             && 2.55 && 30 & 30 & 20 \\
             && 7.65 && 35 & 35 & 20 \\
             \cmidrule{2-7}
                   & \multirow{3}{*}{\includegraphics[width=0.05 \linewidth]{kernel5.PNG}  \includegraphics[width=0.05\linewidth]{kernel7.PNG}} 
             
              & $\sqrt{2}$ && 25 & 25 & 30 \\ 
              && 2.55 && 30 & 30 & 30 \\
              && 7.65 && 35 & 35 & 30 \\

    \midrule
    \multirow{2}{*}{Pixel-wise inpainting } 
    & $0.1$ & 0 &&255 & 1& 118 \\ 
    & $0.2$ & 0 && 255& 1 & 200 \\

            \bottomrule
        \end{tabular}
    }
    
    \label{tab:Parameters PnPADMM}
    \vspace{-8pt}
\end{table}}
\vspace{5pt}


}

\bibliography{references}

\end{document}